\documentclass[a4paper, 11pt, oneside]{article}
\pdfoutput=1
\usepackage{aapreprint}

\usepackage{graphicx,wrapfig,float,slashed}
\usepackage{amsmath,amssymb,dsfont,epsfig,graphicx}
\usepackage{ragged2e}
\usepackage[utf8]{inputenc}
\usepackage{enumitem}
\usepackage{epstopdf}
\usepackage[table]{xcolor}
\usepackage[toc]{appendix}
\usepackage{multirow}
\usepackage[normalem]{ulem}
\usepackage[font={it,small},skip=1pt]{caption}
\allowdisplaybreaks
\usepackage{pifont}
\usepackage{subcaption}
\usepackage{mathtools}

\newcommand{\nc}{\newcommand}
\nc{\non}{\nonumber}
\nc{\hc}{\hbox {H.c.}}
\nc{\noi}{\noindent}
\nc{\barx}{\bar{x}}
\nc{\pbarn}{\;\hbox {pb}}
\nc{\fbarn}{\;\hbox {fb}}

\nc{\hsp}{\hspace{0.5cm}}
\nc{\lsp}{\hspace{1cm}}
\nc{\Lsp}{\hspace{2cm}}
\nc{\LLsp}{\lsp\lsp}
\nc{\lra}{\longrightarrow}
\nc{\p}{\prime}
\nc{\sgn}{\text{sgn}}
\nc{\ph}{\varphi}
\nc{\op}{{\cal O}}
\nc{\eq}{\text{Eq.~}}

\nc{\beq}{\begin{equation}}  \nc{\eeq}{\end{equation}}
\nc{\bea}{\begin{eqnarray}}  \nc{\eea}{\end{eqnarray}}
\nc{\baa}{\begin{array}}     \nc{\eaa}{\end{array}}
\nc{\bit}{\begin{itemize}}   \nc{\eit}{\end{itemize}}
\nc{\ben}{\begin{enumerate}} \nc{\een}{\end{enumerate}}
\nc{\bce}{\begin{center}}    \nc{\ece}{\end{center}}
\nc{\bpm}{\begin{pmatrix}}   \nc{\epm}{\end{pmatrix}}
\nc{\bvt}{\begin{verbatim}}  \nc{\evt}{\end{verbatim}}

\def\lsim{\mathrel{\raise.3ex\hbox{$<$\kern-.75em\lower1ex\hbox{$\sim$}}}}
\def\gsim{\mathrel{\raise.3ex\hbox{$>$\kern-.75em\lower1ex\hbox{$\sim$}}}}

\def\udots{\mathinner{\mkern1mu\raise1pt\vbox{\kern7pt\hbox{.}}\mkern2mu\raise4pt\hbox{.}\mkern2mu\raise7pt\hbox{.}\mkern1mu}}

\def\gev{\;\hbox{GeV}}
\def\tev{\;\hbox{TeV}}

\def\mpl{M_{\text{Pl}}}
\def\mkk{M_{\text{KK}}}
\definecolor{agray}{rgb}{0.95, 0.95, 0.99}
\def\mphi{m_\phi}
\def\mh{m_h}

\def\cth{c_{\theta}}
\def\sth{s_{\!\theta}}
\def\bk{b_{\textsc k}}
\def\cm{c_{\textsc m}}
\def\cw{c_{\textsc w}}
\def\cz{c_{\textsc z}}
\def\cx{c_{\textsc x}}

\def\cf{c_{\psi}}
\def\ch{c_{\textsc h}}
\def\bw{b_{\textsc w}}
\def\bz{b_{\textsc z}}
\def\bx{b_{\textsc x}}
\def\ba{b_{\gamma}}
\def\bg{b_{g}}
\def\baz{b_{\gamma\textsc{z}}}
\def\bh{b_{\textsc{h}}}

\def\uv{\text{\tiny UV}}
\def\ir{\text{\tiny IR}}

\nc{\ctw}{\cos\theta_{\textsc w}}
\nc{\stw}{\sin\theta_{\textsc w}}
\nc{\cwsq}{\cos^2\theta_{\textsc w}}
\nc{\swsq}{\sin^2\theta_{\textsc w}}
\definecolor{cred}{rgb}{0.6, 0.0, 0.0}
\setlist[itemize]{itemsep=0em, topsep=0.3em}

\preprint{\href{https://link.springer.com/article/10.1007/JHEP05(2020)093}{\textcolor{gray}{JHEP\,{\bf05}\,(2020)\,093}}}
\arxivnumber{1912.06645}

\begin{document}
\title{A light dilaton at the LHC}
\author[]{Aqeel Ahmed,}
\author[]{Alberto Mariotti,}
\author[]{and Saereh Najjari}
\affiliation[]{Theoretische Natuurkunde \& IIHE/ELEM,\\ Vrije Universiteit Brussel, Pleinlaan 2, 1050 Brussels, Belgium}
\emailAdd{aqeel.ahmed@vub.be}
\emailAdd{alberto.mariotti@vub.be}
\emailAdd{saereh.najjari@vub.be}

\abstract{
In this paper, we explore the possibility that a light dilaton can be the first sign of new physics at the LHC.
The dilaton could emerge in approximate scale invariant UV completions of the SM as the Goldstone boson associated with the spontaneous breaking of the scale invariance.
We study in detail the phenomenology of the dilaton at the LHC in the mass range of [$10-300$] GeV including the case where the dilaton can mix with the SM Higgs boson, leading to an interesting interplay between direct and indirect constraints. 
A possibility that the dilaton acts as a portal to a dark sector is also considered. As a minimal realization, the dark sector includes a dark photon lighter than the dilaton implying sizeable missing energy signatures.
Several simplified benchmark models that can encode different UV completions are discussed, for which we scrutinize the current and future LHC reach.}


\toccontinuoustrue

\maketitle
\flushbottom

\vspace{-0.7cm}
\section{Introduction}
\label{Introduction}

The absence of new physics at the Large Hadron Collider (LHC) and in other particle physics experiments makes it necessary to revisit the paradigms of beyond the Standard Model~(BSM) physics with the purpose of finding if something could have been overlooked in our strategy to search for new physics. 
In particular, the LHC has strongly constrained many conventional BSM scenarios where new particles significantly coupled with the Standard Model (SM) are predicted around the TeV scale. 
On the other hand, it is also interesting to investigate whether LHC has exploited its maximal constraining power for light new physics which is very weakly coupled to the SM particles. 
For instance, in the case in which such small couplings are determined by higher dimensional operators, it is important to assess the reach of the LHC on the characteristic scale of these couplings and compare it with the direct searches for new resonances expected at such scale.
In this context, several studies have been performed for the case of light pseudo-scalar (axion-like) particles, which are the Goldstone bosons associated with the spontaneous breaking of a global symmetry.

Here we focus on a well motivated scenario for a light scalar degree of freedom, that is the \emph{dilaton}.
The dilaton is a pseudo-Goldstone boson associated with the spontaneous breaking of the scale invariance.
It can typically arise in BSM scenarios involving strongly coupled approximately scale invariant UV completions of the SM (such as the composite Higgs)
and their holographic dual warped extra dimensional models, aiming at addressing the hierarchy problem or other open issues of the SM. 
In the case of warped extra dimensional models, the corresponding light mode is usually referred to as the radion.
In the following, we generically refer to this new light scalar as a dilaton independently of its UV completion origin.
The conventional new physics signatures of this kind of models are usually the top-partners or other new states in the strong sector. 
However, the null results from the LHC direct searches have pushed the mass scale of these states well above a TeV~\cite{Sirunyan:2019xeh}. 
At the same time, in some cases the indirect constraints on these models are even more stringent $\sim\!\op(10)\tev$ and they are mainly driven by the electroweak precision tests (EWPTs) and flavor physics\,\footnote{Note that for instance flavor issues associated with these strongly coupled models could be ameliorated if the strongly coupled theory is approximately conformal invariant in the UV along the lines of Ref.~\cite{Luty:2004ye}.}, see e.g.~\cite{Ahmed:2019zxm}.
Taking the direct and indirect constraints at face value, it appears that the new physics scale associated with the strong dynamics may well be out of the LHC reach. 
On the other hand, if the scale invariance is primarily spontaneously broken (and certain conditions are met, see below), the resulting pseudo-Goldstone boson might be significantly lighter than the other states of the strong sector.
In this perspective, it is interesting to investigate whether a light dilaton can be the first sign of BSM physics at the LHC, assuming that the other new states associated with the strong dynamics are beyond reach.

In addition, it is natural to envision the possibility that the dilaton provides a portal between the SM and a 
dark sector~\cite{Bai:2009ms,Lee:2013bua,Blum:2014jca,Efrati:2014aea,Kim:2016jbz,Ahmed:2019kgl}. 
In this work, we assume a strongly coupled approximate scale invariant dark sector which contains a relatively light vector boson ({\it dark photon}) of a dark $U(1)_X$ gauge symmetry. 
The other dark sector states are assumed to be 
heavier than the dilaton and hence play no significant role in the phenomenology of the dilaton.
We focus on regimes where the dark photon mass is smaller than the dilaton, such that the dilaton can decay to dark photons.
The invisible decay of the dilaton gives rise to 
missing energy signatures at the LHC, which can be constrained by mono-jet searches.
Generically, one could investigate if the dark sector includes a viable dark matter (DM) candidate which constitutes the observed DM relic density~\cite{Bai:2009ms,Lee:2013bua,Blum:2014jca,Efrati:2014aea,Kim:2016jbz}. 
In this paper our focus is on the LHC phenomenology of the light dilaton, and we leave the detailed DM phenomenology for future works.

It is relevant to comment under which conditions on the UV completion one could expect the appearance of a light dilaton in the spectrum, possibly parametrically lighter than the characteristic scale~$f$ of the underlying symmetry breaking, where other new particles are generically present. 
For a light dilaton, a possible mechanism has been suggested by Contino-Pomarol-Rattazzi (CPR)~\cite{CPR:2010} and further elaborated in~\cite{Chacko:2012sy, Bellazzini:2012vz, Bellazzini:2013fga, Coradeschi:2013gda,Cox:2014zea,Chacko:2014pqa,Megias:2014iwa,Abu-Ajamieh:2017khi}. 
The key ingredient of this construction is the explicit breaking of the scale-invariance by an almost marginal operator, which induces a slow running (small beta function) 
of the quartic coupling in the dilaton potential. 
In~\cite{CPR:2010,Bellazzini:2012vz,Coradeschi:2013gda} an explicit realization in 5D warped Randall-Sundrum (RS)-like scenario~\cite{Randall:1999ee} with Goldberger-Wise (GW) stabilization mechanism~\cite{Goldberger:1999uk} were provided, where the mass of the dilaton can indeed be tuned to be smaller than the size of the extra dimension\,\footnote{The existence of a light dilaton in holographic models at the conformal transition has been also investigated recently in~\cite{Pomarol:2019aae}.}.
These recent developments further motivate the phenomenological study of a light dilaton at the LHC.

Hence, in this paper we adopt a bottom-up approach and study the phenomenology of an effective theory of a light dilaton, assuming that all the other BSM particles are out of the LHC reach. 
The structure of the low energy effective action is then determined by the nonlinearly realized scale invariance below the scale~$f$. 
In particular, the dilaton couplings to the SM are induced through higher dimensional operators suppressed by the scale~$f$.
However, the values of the Wilson coefficients depend on the specific UV completion and details of the scale symmetry breaking.
In addition, also the mixing of the dilaton with the Higgs is, in general, a model dependent feature.
Hence, in the following for concreteness, we will focus on a few benchmark models for our quantitative analysis.

We will consider three scenarios for the phenomenological study, without and with the portal to the dark sector:
\begin{itemize}
\item[(i)] \emph{Minimal dilaton}, where the dilaton mixes with the SM Higgs via mass mixing and apart from that it couples to the SM via the trace of the energy-momentum tensor.
\item[(ii)] \emph{Holographic dilaton}, where the dilaton and the SM Higgs mix through a kinetic term. Moreover, in this model we assume the partial composite framework~\cite{Kaplan:1991dc} where the Higgs doublet and right-handed top quark are composite states, while all the other SM fields are elementary. This a holographic realization of a 5D warped extra dimensional RS-like model.
\item[(iii)] \emph{Gauge-philic dilaton}, where the dilaton couples only to the field strength tensors of the gauge bosons due to the running of the gauge couplings, and not to any of the mass terms. 
This is an extreme simplification, since one expects that in this case the SM masses will backreact on the dilaton potential. However we consider it for phenomenological purposes in order to illustrate the collider reach on such elusive scenario.
\end{itemize}
We focus on the mass window of $[10\!-\!300]\gev$ for the dilaton, which has not yet been analyzed thoroughly in previous studies (see however \cite{Barger:2011nu, Cox:2013rva, Blum:2014jca, Jung:2014zga, Kim:2016jbz, Sachdeva:2019hvk} for the existing studies).
Indeed, the case of very light dilaton masses below $10\gev$ has been recently investigated in~\cite{Abu-Ajamieh:2017khi}, while the high mass region has been investigated in several works~\cite{Csaki:2007ns,Foot:2007as,Goldberger:2008zz,Fan:2008jk,Vecchi:2010gj,Appelquist:2010gy, Barger:2011hu,Hur:2011sv,Coriano:2012nm, Chacko:2012vm,Bellazzini:2012vz,Abe:2012eu,Cao:2013cfa,Megias:2014iwa,Bellazzini:2015nxw,Ahmed:2015uqt,Bandyopadhyay:2016fad,Chakraborty:2017lxp}. 
Our goal here is to provide broad coverage of the phenomenology of a light dilaton at the LHC, that could be easily re-interpreted in diverse UV completions.
Furthermore, in our analysis we will identify regions of parameter space where the existence of a light dilaton is compatible with LHC exclusion limits and where the hierarchy between the dilaton mass 
$m_\phi$ and the scale $f$ of the spontaneous breaking of the scale invariance is still moderate, i.e.~$m_{\phi}/f \!\sim \!\op(1\!-\!10\%)$. 
These represent promising physics cases that dedicated LHC searches for light new states in the future LHC run will be able to further explore and test.

The paper is organized as follows. In Sec.~\ref{sec:theory}, we outline the low energy effective theory where only a light dilaton (and possibly a dark photon) are present in addition to the SM. 
We assume that all the other new particles associated with the UV theory to be beyond the LHC reach. 
In the effective action we consider the most general interactions allowed by the nonlinearly realized scale symmetry up to dimensions five operators, 
and we also
include non-trivial dilaton-Higgs mixing. Furthermore, three well motivated scenarios are introduced.
The detailed collider phenomenological analysis of a light dilaton in the mass range $[10\!-\!300]\gev$, without and with a dilaton portal to the dark sector, is then performed in Sec.~\ref{sec:pheno}.
We conclude our work in Sec.~\ref{sec:con}. 
The supplementary material including the Feynman rules is given in Appendix~\ref{sec:feyn_rules}.

\section{Effective theory for a light dilaton}
\label{sec:theory}
In the following we adopt a bottom-up approach such that the low-energy effective theory contains  only the dilaton as a light degree of freedom in addition to the SM states
(later on we will include also a dark photon).
The UV theory is assumed to be strongly coupled with approximate scale invariance which is broken spontaneously at a scale $f$.
The rest of new physics associated with such strong dynamics is taken to be at mass scale $m_\ast = g_\ast f$ and beyond the reach of the LHC, 
where $1\lesssim g_\ast\lesssim 4\pi$ is a generic strong coupling.
In the phenomenological analysis we will comment on the regions of validity for this assumption for each benchmark case studied.

In the effective theory the scale invariance is nonlinearly realized such that the dilaton is embedded in a conformal compensator field defined as $\chi\!\equiv\!f e^{\phi_0/f}$, where $\phi_0$ is the dilaton fluctuation and $f$ is the vacuum expectation (VEV) of $\chi$. 
Note that under the scale transformation $x^\mu\to x^{\prime\mu}\!=\!e^{-\lambda} x^\mu $ and  $\chi(x)\to\chi^\p(x^\p)\!=\!e^{\lambda}\chi(x)$.
The dilaton coupling with the SM can be deduced by inserting appropriate powers of the compensator field in the SM Lagrangian to make it
scale invariant~\cite{Chacko:2012sy,Bellazzini:2012vz}.
In particular, there are dimension five operators suppressed by the scale $f$ induce the dilaton couplings with the SM fields.
However, while the Lorentz structure of such couplings is given, the precise value of the Wilson coefficients depends on the UV completion.
Hence, in our study we employ the effective Lagrangian with generic dimension five couplings of the dilaton to the SM as\,
\begin{align}
{\cal L}^{{\rm int}}_{\text{eff}}\!&=\frac{\phi_0}{f}\bigg[\!\bh\partial_\mu h_0\partial^\mu h_0-2\ch m_{h_0}^2h_0^2 - \cf^i m_{\psi_i} \bar \psi_i\psi_i +2\cw m_{W}^2 W^{+}_\mu W^{-\mu} +\cz m_{Z}^2 Z_\mu Z^\mu     \notag\\
&\quad+\!\frac{\alpha_\textsc{em}}{8\pi}\!\Big(\!\ba F_{\mu\nu}F^{\mu\nu}\!+\!2\bw W_{\mu\nu}^+ W^{-\mu\nu} \!+\!\bz Z_{\mu\nu}Z^{\mu\nu}\!+\!2\baz F_{\mu\nu}Z^{\mu\nu}\! \Big)\!+\!\frac{\alpha_{s}}{8\pi}\bg G_{\mu\nu}^a G^{a\mu\nu}\!\bigg],		\label{eq:sint}
\end{align}
where the constants $b$'s and $c$'s are $\sim\!\op(1)$ model dependent parameters that we will specify in the following for the benchmark models considered. 
The coefficients $c$'s parameterize the dilaton couplings with the mass terms in the SM and they are equal to unity to respect the nonlinearly realized scale invariance. 
Deviation from unity, i.e. $c\!\neq\!1$, captures possible explicit scale symmetry breaking effects, including anomalous dimensions for fermions.

In the effective action~\eqref{eq:sint} the coefficients $b$'s parametrize the dilaton interactions with the field strength of the gauge bosons and are defined as the coefficients of the $\beta$-function, i.e. $\beta(g)\!=\!-b\, g^3/(16\pi^2)$.
These coefficients are model dependent and generically get UV and IR contributions due to the running of the gauge coupling above or below the scale $\sim \!4\pi f$, denoted as $b^{\uv}$ and $b^{\ir}$, respectively.
The effective couplings in~\eqref{eq:sint} are the difference of the IR and UV contributions to the $\beta$-function, i.e. $b_i \!\equiv\!b_i^{\ir}-b_i^{\uv}$.
We refer to~\cite{Chacko:2012sy,Bellazzini:2012vz} for a detailed discussion of these terms.
Note that the effective interactions of the dilaton with the massless gauge bosons receive important one-loop corrections involving the dilaton coupling with the massive fermions and gauge bosons.
For low dilaton mass these loop effects partially cancel with the contribution of the $\beta$-function coefficients $b$'s, in agreement with the consistent decoupling of heavy states.
Without an exact description of the UV dynamics the UV contributions to the $\beta$-function coefficients $b^{\uv}$ are essentially free parameters.
A standard scenario is when the UV contributions are assumed to be vanishing or negligible (i.e. $b^{\uv}\!=\!0$) and the $b$ constants are given by the running of the IR (SM) states only.
In this case the IR $\beta$-function coefficients $b^\ir$ are the ones of the SM, i.e. $b_3^\ir\!=\!7,\, b_2^\ir\!=\!19/6$, and $b_1^\ir\!=\!-41/6$ for the $SU(3)_\textsc{qcd}$, $SU(2)_L$, and $U(1)_Y$ gauge groups, respectively.
Instead, in one of the benchmark studied, we will consider a different set of coefficients $b^{\uv}$ taking inspiration from the partial composite framework with a holographic realization where the SM fields are embedded in an extra dimensional warped scenario~(see Section~\ref{sec:holodil}).

Besides the above interaction Lagrangian linear in the dilaton field, there are higher order interactions involving more that one dilaton fields and the SM fields. Such terms are significantly model dependent
and we do not discuss them here.
However there is another possible source of interaction which is due to the dilaton mixing with the SM Higgs. 
Such a mixing can be generated via dilaton-Higgs kinetic and/or mass mixing. 
These depend on the specific embedding of the SM Higgs in the sector responsible for the breaking of the scale invariance.
Without specifying the details of the electroweak symmetry breaking,
we remain here agnostic about the nature of  the SM Higgs field.
We then consider the following Lagrangian for the dilaton-Higgs system after the electroweak symmetry breaking (EWSB), up to term quadratic in the fields,
\begin{align}
{\cal L}^{(2)}_{\text{eff}}=\frac12 \partial_\mu h_0\partial^\mu h_0+\frac{1}{2} \partial_\mu \phi_0\partial^\mu\phi_0-\frac12 m_{h_0}^2 h_0^2-\frac12 m_{\phi_0}^2\phi_0^2 +\bk\partial_\mu h_0\partial^\mu \phi_0+\cm m_{\phi_0}^2h_0 \phi_0, 	\label{eff_lagrangian}
\end{align}
where $h_0$ is the SM-Higgs scalar,
$m_{h_0}\!\equiv\! \sqrt{2\lambda_0}\,v$ is the bare Higgs mass, and $v\!=\!246\gev$ is the SM Higgs VEV, while $\phi_0$ denotes the dilaton in the interaction basis.
Note that the mass terms involving the dilaton/Higgs boson represent explicit scale symmetry breaking operators.
The last two terms in Eq.~\eqref{eff_lagrangian} introduce kinetic and mass mixings between the dilaton state and the SM Higgs, parameterized by dimensionless constants $\bk$ and $\cm$, respectively. 
We assume that such coefficients are $\bk,\cm\lsim\!\op(1)$. 
Note however that we parameterize the dilaton-Higgs mass mixing term with the dilaton mass $m_{\phi_0}^2$ (times a dimensionless parameter~$\cm$). Given that such a term is a source of explicit breaking of the scale invariance, one could expect it to be proportional to the dilaton mass.
A possible source of dilaton-Higgs kinetic mixing could be the gauge invariant dimension four operator $|H|^2\widehat {\cal R}$, where $\widehat {\cal R}$ is the Ricci scalar,
as we will explain here below.

In our phenomenological study we also include trilinear couplings between the SM-like Higgs and the dilaton, 
since they typically induce 2-body decay modes and can affect the SM-like Higgs properties or the dilaton branching ratios (depending on the mass regime).
We neglect instead higher order corrections, 
assuming that they will not affect significantly the phenomenology.
The trilinear dilaton-Higgs interactions can emerge, after the rotation to the mass eigenstate, from the following three sources:
\bit
\item[(a)] The dilaton coupling from the effective interaction Lagrangian~\eqref{eq:sint}, 
\item[(b)] The trilinear coupling in the Higgs potential, i.e.
\beq
\mathcal{L} \supset -\frac{1}{2}\frac{m_{h_0}^2}{v}h_0^3.	\label{eq:tri_h}
\eeq
\item[(c)] Additional dilaton-Higgs trilinear terms may arise if a non-minimal coupling of the SM Higgs with the Ricci scalar is assumed, i.e.\
\begin{align}
\mathcal{L} \supset \sqrt{\!-\widehat g}\,\xi\, |H|^2\widehat {\cal R} &=  -6 \xi\frac{v}{f}\bigg(1+\frac12 \frac{h_0}{v}\bigg) h_{0} \square \phi_{0}+\dots, \label{eq:tri_xi}
\end{align}
where $\xi$ parameterizes the Higgs-gravity non-minimal coupling and ellipses denote terms suppressed by $f^2$ or more. 
\eit
Above $\widehat g$ is the determinant of the 4D metric $\widehat g_{\mu\nu}\!=\!\eta_{\mu\nu}\chi^{2}\!/\!f^2$, where $\eta_{\mu\nu}$ is the Minkowski background metric. The 4D Ricci scalar $\widehat{\cal R}$ is constructed out of the metric $\widehat g_{\mu\nu}$ which contains the dilaton field as Weyl transformation.
The quadratic operator in~\eqref{eq:tri_xi} induces the kinetic mixing in~\eqref{eff_lagrangian} . 
The trilinear operator can be mapped to additional contributions to the coefficients of operators already present in the effective interaction Lagrangian~\eqref{eq:sint} by using the lowest order Higgs equation of motion.
We neglect the dilaton self-interacting terms of the form $\phi_0^3$ from the dilaton potential, which are expected to be proportional to $m_{\phi_0}/f$, and hence negligible in the mass regime we are interested in. 
Note that with these assumptions the Higgs decay mode into a pair of dilatons is absent if the mixing parameters are vanishing (see Appendix \ref{sec:feyn_rules} for the explicit formulas).

In Tab.~\ref{tab:coeff_models}, we collect the exact values of the coefficients specifying the effective Lagrangian for the concrete models that we will consider in the following. 
The details of each benchmark scenario are described in subsection~\ref{sec:bsm_models}.
As mentioned, from the effective field theory perspective the dilaton coupling to the SM fields are determined by the structure of nonlinearly realized scale invariance.
However, since the scale invariance is explicitly broken by operators in the IR and UV, there are generically 
modifications to the dilaton couplings induced in the low energy theory.
The same explicit breakings are also responsible to generate the non-zero mass for the dilaton field, therefore at leading order the corrections to the dilaton couplings are proportional to $\sim\! m_\phi^2/f^2$. 
Such corrections would then be typically small for the dilaton masses we are interested in, i.e. $\!m_\phi^2/f^2\!\ll\!1$.

Within this parameterization and assumptions, we can proceed in rotating the system to the mass eigenstates which will be used for the study of the considered benchmarks.
In particular, the kinetic and mass mixings of the dilaton and SM Higgs in~\eqref{eff_lagrangian} can be removed by the following transformation into the mass eigenstates ($\phi,h$),
\begin{align}
\bpm \phi_0\\ h_0\epm=\frac{1}{Z}\!\!\bpm \cth&&-\sth\\ Z\sth\!-\!\bk\cth&&Z\cth\!+\!\bk\sth\epm \bpm \phi \\ h\epm,   \label{h-r_mixing}
\end{align}
where $Z\!\equiv\! \sqrt{1-\bk^2}\,, \cth\equiv\cos \theta\,, \sth\equiv\sin \theta\,$, and the mixing angle $\theta$ is defined as,
\beq
\tan2\theta=-\frac{2\sqrt{1-\bk^2}\big(\cm m_{\phi_0}^2+\bk m_{h_0}^2\big)}{\big(1+2\bk\cm\big)m_{\phi_0}^2-\big(1-2\bk^2\big)m_{h_0}^2}~.  \label{tan2theta}
\eeq
The physical mass-eigenvalues for the states $h$ and $\phi$ are, 
\begin{align}
m_{h/\phi}^2&=\frac{1}{2(1- \bk ^2)}\bigg[m_{h_0}^2 +m_{\phi_0}^2+2  \bk \cm m_{\phi_0}^2 	\notag\\
& \quad\qquad\mp \!\sgn(m_{\phi_0}\!-\!m_{h_0}) \sqrt{\big(m_{\phi_0}^2-m_{h_0}^2\big)^2+4 m_{\phi_0}^2  (\bk+\cm)\big(\cm m_{\phi_0}^2+\bk m_{h_0}^2 \big)}\bigg].	
\end{align}
In the following, we fix the SM-like Higgs physical mass $m_h\!=\!125\gev$, while the dilaton physical mass is taken in the range~$\mphi\!\in\![10\!-\!300]\gev$. 
The above mass relations along with Eq.~\eqref{tan2theta} can be solved to fix the Higgs and dilaton bare masses $m_{h_0}^2$ and $m_{\phi_0}^2$, respectively. 
Note that the region of the parameter space which leads to the square of the bare masses negative, i.e. $m_{h_0/\phi_0}^2\!<\!0$, would be referred to as an unphysical region, since it would lead to unstable vacuum configuration.

In the following we will study the phenomenology of concrete models outlined in the subsections below.
The relevant signatures will involve direct LHC searches for the mass eigenstate $\phi$ (mostly dilaton) and its decay products, as well as indirect constraints arising from
modification of the Higgs coupling induced by the mixing.
We will present our results in the $m_{\phi}$ vs $f$ plane to show the current and future coverage of the LHC on these type of models. 
As we discussed in the Introduction, realizing a light dilaton (so with $m_{\phi} \ll f$) could require some tuning 
and/or specific conditions on the dilaton potential.
So, in our phenomenological analysis we will display an indicative line $m_{\phi}/f = 1\%$ to divide the parameter space in two regions. 
The region with a smaller ratio of $m_{\phi}/f$ should be considered \emph{fine-tuned} from the theory perspective. 
On the other hand, the interesting question that we aim to answer is how much of the complementary parameter space with $m_{\phi}/f > 1\%$ 
can be covered by LHC at present and in the future searches.

\subsection{Dilaton portal to a dark sector}
\label{sec:dark}
In this subsection, we extend the scenario discussed above by considering the possibility that the dilaton provides the portal to the dark sector which may  include a DM candidate. 
We assume that only one state in the dark sector is light 
and should be included in the effective theory, whereas the other dark sector states are heavy and 
do not play any significant role in the dilaton phenomenology.
One of the simplest possibility is that the dark sector employs an Abelian gauge symmetry $U(1)_X$ and the corresponding gauge boson $X_\mu$, which we refer to as the {\it dark photon},
is the only light degree of freedom of the dark sector.
 The coupling of the dilaton with the dark gauge bosons are dictated by nonlinearly realized scale invariance, and 
the dilaton would act as a portal from the SM to the dark sector.

Note that the dark photon can be stable due to the dark $U(1)_X$ charge conjugation symmetry ${\cal C}$ under which it transforms as, $X_\mu\overset{{\cal C}}{\to} -X_\mu$ (see e.g.~\cite{Ahmed:2017dbb}), and hence could provide a viable dark matter candidate~\cite{Blum:2014jca}.
As mentioned above, we are interested in possible collider signatures of such scenario and hence we focus on the regime
where the dark photon mass $m_X\lesssim m_\phi/2$, such that the dilaton can decay to dark photons.
This can be probed at the LHC via the mono-jet searches, which we discuss in the next section.

We describe the dynamics of the dark photon in the low energy theory by the following simplified Lagrangian 
\beq
{\cal L}_{\rm dark}=-\frac14 X_{\mu\nu} X^{\mu\nu} +\frac12 m_X^2 X_\mu X^\mu +\frac{\phi_0}{f}\Big[\cx m_{X}^2 X_\mu X^\mu +\frac{\alpha_\textsc{x}}{8\pi}\bx\, X_{\mu\nu}X^{\mu\nu}\Big],		\label{eq:Ldark}
\eeq
where the portal couplings (the last two terms) arise by inserting the spurion field $\chi$ in order to nonlinearly realize scale invariance in the dark photon Lagrangian.
In particular, $\alpha_\textsc{x}$ is the dark sector fine-structure constant, defined as $\alpha_\textsc{x}\!\equiv\! g_\textsc{x}^2/(4\pi)$, where $g_\textsc{x}$ is the dark $U(1)_X$ gauge coupling, which we assume to be $\op(1)$. 
The coefficient $\bx$ captures the running effects of the dark gauge coupling $g_\textsc{x}$, which we assume to be large~$\op(10)$, as it can be for instance realized if the dark sector involves large number of states charged under the dark $U(1)_X$ gauge symmetry. 
The constant $\cx$ measures possible explicit breaking effects of the scale invariance.

In the following phenomenological study we fix the dark photon mass $m_X\!=\!1\gev$, however, our results are fairly independent of the dark photon mass as long as $m_X\!\lesssim\!1\gev$ (see Eq.~\eqref{eq:Gamphixx}). 
In this perspective one can think that the dark photon mass is a free parameter (given $m_X\!\lesssim\!1\gev$) which, for instance, can be fixed by requiring that it reproduces the correct DM relic abundance. 
We leave a detailed study about possible mechanisms that could lead to the correct relic abundance of this dark matter candidate for future studies.
Here we focus on the collider signatures of a possible dark sector decay of the dilaton, with the purpose of providing results which could be interpreted in dilaton portal DM models.

\subsection{BSM benchmark models}
\label{sec:bsm_models}
For concreteness in the following we consider three different classes of BSM models with a light dilaton of which we study the phenomenology at the LHC. 

\subsubsection{Minimal dilaton model}

In the first scenario we make the simplifying assumption that there is no kinetic mixing, i.e. $\bk\!=\!0$.
We allow however for a dilaton-Higgs mass mixing which is parameterized by the coefficient $\cm$ in Eq.~\eqref{eff_lagrangian}. In particular we study two cases, 
$\cm = 0$ (no mixing) and $\cm= 0.1$.
We label this scenario the \emph{minimal dilaton} model, since it represents the simplest scenario for the dilaton-Higgs mixing.
Indeed, the mixing structure in the scalar sector is analogous to the singlet scalar extensions of the SM.
However, the dilaton possesses direct couplings with the SM fields encoded in the dimension-five operators
suppressed by the scale $f$ as in Eq.~\eqref{eq:sint}.
In particular, it has couplings with the fermions and massive gauge bosons proportional to their masses, and to the massless gauge bosons proportional to the coefficients of the $\beta$-functions of the gauge couplings. 
The minimal dilaton model can be realized as a low energy theory of a strongly coupled nearly scale invariant UV complete theory, see e.g.~\cite{Luty:2004ye,Appelquist:2010gy,Hur:2011sv}.

The mass mixing between the dilaton and Higgs are removed by the rotation matrix which is orthogonal and unitary. The rotation matrix~\eqref{h-r_mixing} with $\bk\!=\!0$ takes the usual form, 
\begin{align}
\bpm \phi_0\\ h_0\epm&=\bpm \cos\theta&&-\sin\theta\\ \sin\theta&&\phantom{-}\cos\theta\epm \bpm \phi \\ h\epm,  \label{eq:massmixing}
\end{align}
where $\theta$ is the mixing angle given as, 
\begin{align}
\tan2\theta&=\frac{\cm \Big(m_h^2+\mphi ^2+\sgn(m_{\phi}\!-\!m_{h})\sqrt{\big(\mphi ^2-m_h^2\big)^2-4 \cm^2 m_h^2 \mphi ^2}\Big)}{\big(m_h^2+\mphi ^2\big)\cm^2 -\sgn(m_{\phi}\!-\!m_{h})\sqrt{\big(\mphi ^2-m_h^2\big)^2-4 \cm^2 m_h^2 \mphi ^2}}\,,	\notag\\
&\approx\begin{dcases}
~2 \cm \Big(\frac{m_{\phi }^2}{m_h^2}\Big)	& \qquad{\rm for}\quad m_{\phi }^2\!\ll\!m_h^2	\\
\!\!-2 \cm \Big(1+\frac{m_h^2}{m_{\phi }^2}\Big)	 & \qquad{\rm for}\quad m_{\phi }^2\!\gg\!m_h^2 
\end{dcases}\,.  \label{eq:thetamm}
\end{align}
All the other parameters of the effective interaction Lagrangian~\eqref{eq:sint} are collected in Tab.~\ref{tab:coeff_models}. The remaining model dependent parameters are the $\beta$-function coefficients for the gauge couplings for which we assume that the UV contribution is negligible and the IR contribution is the one of the SM. 
Hence the explicit values of $b_i$-coefficients are $b_3\!=\!7,\, b_2\!=\!19/6$ and $b_1\!=\!-41/6$. 
Such a choice of $b_i$-coefficients can be realized, for instance, in strongly coupled nearly scale invariant composite Higgs models where all the SM fields are composite.
Moreover, in this scenario when we include the possibility that the dilaton act as a portal to the dark sector we will consider the coupling via the mass term 
but not the coupling through the RG running, that is $\cx\!=\!1$ and $\bx\!=\!0$ in Eq.~\eqref{eq:Ldark}.

Finally, the trilinear interactions of the dilaton-Higgs fields in the minimal dilaton scenario has two sources, the dimension five interactions Eqs.~\eqref{eq:sint} and the trilinear of the Higgs \eqref{eq:tri_h}.

\subsubsection{Holographic dilaton model}
\label{sec:holodil}
The second scenario we consider is a light dilaton in a holographic model realized in a 5D warped extra dimensional RS-like scenario~\cite{Randall:1999ee}. 
The RS-like scenario involves one extra dimensions with an $S_1/\mathbb{Z}_2$ orbifold and two D3-branes located at the fixed point of the orbifold action,
respectively the IR and the UV branes.
The five dimensional metric can be parameterized as:
\beq
ds^2=e^{-2k |y|}\eta_{\mu\nu}dx^\mu dx^\nu-dy^2, \label{rs_metric}
\eeq
where $k$ is the  curvature of the 5D geometry, $0\leq y\leq \pi R$ is the extra dimensional coordinate, and $R$ is the size (radius) of  the fifth dimension. 
In order to solve the gauge hierarchy problem, 
one typically requires $kR\!\sim \!\op(10)$ in such RS-like models. 
The fluctuation corresponding to the inter-brane distance is referred to as the radion,
and plays the role of the dilaton in the holographic 4D effective theory~\cite{Rattazzi:2000hs}.
The interbrane distance is stabilized through the Goldberger-Wise mechanism which also provides the dilaton/radion a mass \cite{Goldberger:1999uk,DeWolfe:1999cp,Csaki:2000zn}.  
The resulting effective action that we consider has been derived in Refs.~\cite{Giudice:2000av,Dominici:2002jv,Dominici:2002np,Grzadkowski:2012ng}.

In the literature, there are many variants of the RS model and the dilaton/radion dynamics depends on the model details. 
We assume that a mechanism like CPR~\cite{CPR:2010, Bellazzini:2013fga, Coradeschi:2013gda} is at work such that a light dilaton can be realized.
We also follow the partial composite paradigm such that the Higgs doublet $H$ and the right-handed top quark $t_R$ are composite states, i.e. localized on the IR brane, while the remaining SM fields (including especially the gauge bosons) are mostly elementary, i.e. localized towards the UV brane or in the bulk~\cite{Csaki:2007ns,Grzadkowski:2012ng,Toharia:2008tm,Chacko:2014pqa,Ahmed:2015uqt}.
Finally we further assume a bulk custodial symmetry, such that the new physics resonances (KK-modes) can be at moderately low scale without conflicting with the EWPTs, in particular the T-parameter~\cite{Agashe:2003zs}.

As already introduced, we also allow in this scenario for a non-minimal dilaton-Higgs mixing resulting from the Higgs coupling with the brane induced Ricci scalar $\widehat{\cal R}$ at the IR brane, parametrized as $\xi |H|^2\widehat{\cal R}$ in \eqref{eq:tri_xi}. In particular, this term includes a kinetic mixing between the Higgs boson and the dilaton, such that $\bk\!=\!6\xi v\!/\!f$.
We collect all the parameters of the effective Lagrangian~\eqref{eff_lagrangian} and the dilaton interactions~\eqref{eq:sint} within this holographic model in Tab.~\ref{tab:coeff_models}, where a notable non-standard parameters are the gauge boson couplings due to their presence in the bulk~\cite{Csaki:2007ns,Chacko:2014pqa,Ahmed:2015uqt}. 
The parameter $c_{\textsc v}\!=\!1-\tfrac{3\pi k R \, m_V^2}{f^2(k/\mpl)^2}$ (where $V\!=\!W,Z$) deviates from 1 proportional to $m_V^2/f^2$ and the bulk volume factor $\pi kR$. We neglect however other possible explicit breaking corrections proportional to $m_\phi^2/f^2$ which could be of the same order.
Furthermore, due to the EWSB on the IR-brane, there are corrections to the flat bulk profiles for the massive gauge bosons which induce corrections to their couplings with the dilaton as well.

The complementary new physics signatures in extra-dimensional models are the searches of the lightest KK states, whose masses are expected to be close to $f$. 
In our analysis, we would like to focus on a regime for $f$ for which these states are beyond the reach of the LHC, such that the light dilaton is in fact the only expected sign of new physics.
In the RS model we can easily map the scale of the KK modes with the scale $f$.
First, the dilaton VEV is related to the geometry of the extra dimension as
\beq
f= \sqrt6 \,\mpl\, e^{-\pi k R}\,.	\label{eq:frs}
\eeq 
Here $\mpl$ is the 4D Planck mass, which is related to the 5D fundamental parameters ($M_5, k, R$) as, 
\beq
\mpl^2=\frac{M_5^3}{k}\Big[1-e^{-2\pi kR}\Big].
\eeq
It is convenient to define the KK scale $\mkk$ which represents the general mass scale of KK states associated with the bulk fields as,
\beq
\mkk\equiv 2\,k \,e^{-\pi k R}, \lsp{\rm such~that}\lsp  f=\sqrt{\frac32}\frac{\mpl}{k}\mkk.	\label{eq:frs}
\eeq 
As already mentioned, we take $kR\!\sim\!10$ having in mind RS realization which can solve the electroweak hierarchy problem. 
Moreover, we take $\mkk \!=\! 4\tev $ which makes the KK resonances approximately out of the LHC reach\,\footnote{Note that physical KK resonances are heavier than the KK scale $\mkk\equiv 2ke^{-\pi kR}$. For instance, the mass of first KK gluon is $m_1^g\simeq 1.2\mkk$ and the mass of the first KK graviton state is $m_1^G\simeq 1.8\mkk$.}.
The value of the curvature scale $k$  in the RS model is a free parameter of order $\mpl$. However, its maximum value is assumed to be $k\!/\!\mpl \simeq 3$.
Indeed, above $k\!/\!\mpl\simeq 3$ higher curvature/loop corrections to the 5D action become relevant \cite{Agashe:2007zd} and therefore, from the theory perspective, that region of parameter space is not robust. 
Hence, for a fixed value of $\mkk = 4\tev $ and requiring $k\!/\!\mpl \leq 3$ we get from Eq.~\eqref{eq:frs} a lower bound on $f$, i.e. $f\gtrsim 1.6$ TeV,  that marks the theoretically motivated region
of our parameter space. We will explicitly display such limit on our final plots where we show the LHC constraints.

As mentioned in Section \ref{sec:theory}, the couplings of the dilaton to the massless gauge bosons, which are induced by the running of the gauge couplings, are 
generically model dependent, and receive UV and IR contributions.
The IR contributions $b_i^{\ir}$ are due to all the low energy states, composite as well as elementary, i.e. the full SM degrees of freedom. 
Instead, all the elementary fields (localized on the UV brane or in the bulk) as well the CFT operators, which are essentially unknown, contribute to the $b_i^{\uv}=b_{\textsc{elem}}^{\uv}+b_{\textsc{cft}}^{\uv}$. 
In the following we assume the UV contributions from the CFT/strong dynamics are vanishing, i.e. $b_\textsc{cft}^{\uv}=0$. Hence, the $b_i^{\uv}$ include only contributions due to the SM elementary fields. Note that all the SM fields except the Higgs doublet $H$ and right-handed top quark $t_R$ are localized on the UV brane or in the bulk. Thus in this case 
the total $\beta$-function coefficients $b_i=b_i^{\ir}-b_i^{\uv}$ are only induced by the composite states, i.e. fields localized on the IR brane:
\beq
b_3=-\frac13\,, 		\lsp  b_2=-\frac{1}{6}\,, 		\lsp  b_1=-\frac{19}{18}\,.		\label{eq:caseB}
\eeq
We will see that the LHC phenomenology of the light dilaton is strongly dependent on the choice of these values of $b$-coefficients since they determine the gluon fusion production cross section\,\footnote{Note that the contributions from bulk gauge kinetic terms proportional to the volume factor $1/kR$ are not included in the $\beta$-function coefficients.}.

Finally, also for this benchmark we consider the possibility of coupling the dilaton to a dark sector.
In order to realize the dark photon scenario as in Section \ref{sec:dark},
we employ a dark $U(1)_X$ bulk gauge symmetry such that the zero-mode is $X_\mu$ with mass $m_X$, while the higher KK modes are at scale of order $\mkk$ and decoupled. 
Furthermore, for simplicity we assume that the dark vector couples to the dilaton mainly through the mass term, whereas its coupling due to the running of dark $U(1)_X$ gauge coupling are negligible. 
More precisely we take $\cx\!=\!1$, $\bx\!=\!0$, and $m_X$ is treated as a free parameter, but we restrict to values $m_X\!\lesssim\!m_\phi/2$.

\subsubsection{Gauge-philic dilaton model}
The last benchmark that we consider covers a class of models where the dilaton couples only to the field strength tensors of the gauge bosons via the running of gauge couplings,
and not to any mass term (including Higgs as well as massive gauge bosons and fermions).
We also assume that there is no dilaton-Higgs mixing, i.e. $\bk\!=\!\cm\!=\!0$.
The interaction Lagrangian takes hence the following simple form
\begin{align}
{\cal L}^{\rm int}_{\text{eff}}\!=\!\frac{\phi}{f}\bigg[\frac{\alpha_\textsc{em}}{8\pi}\Big(\ba F_{\mu\nu}F^{\mu\nu}+2\bw W_{\mu\nu}^+ W^{-\mu\nu} +\bz Z_{\mu\nu}Z^{\mu\nu}+2\baz F_{\mu\nu}Z^{\mu\nu} \Big)+\frac{\alpha_{s}}{8\pi}\bg G_{\mu\nu}^a G^{a\mu\nu}\bigg], \label{eq:lag_anom}
\end{align}
where the explicit form of $b$ coefficients are collected in Tab.~\ref{tab:coeff_models}. 
As already explained, the $b$ parameters are the coefficients of the $\beta$-functions of the SM gauge groups.
We consider that there is no contribution from the UV physics to the gauge $\beta$-functions, such that $b_i^{\rm UV}=0$ and $b_i=b_i^{\rm IR}$, where $b_i^{\rm IR}$ only contains the contributions from the SM. Hence the value of coefficients of the beta functions are $b_3\!=\!7,\, b_2\!=\!19/6$ and $b_1\!=\!-41/6$. 

Note that in this extreme case
the mass parameters of the SM are sources of explicit breaking of the scale invariance which do not respect the nonlinearly realized symmetry.
This scenario is considered mainly for phenomenological interest to explore the reach of the LHC in this extreme regime
where the dilaton does not couple to any source of mass terms\,\footnote{
In principle, one can imagine a possibility where the UV theory is close to a conformal point such that all the interactions proportional to mass terms vanish~\cite{Giudice:2000av}.}.

Also for this benchmark, we consider in addition the possibility that the dilaton is a portal to a dark sector. 
In this case, the dilaton also couples to a dark vector due to the running of dark $U(1)_X$ gauge coupling and we assume consistently that its coupling with the mass term is zero, i.e.
we consider the Lagrangian~\eqref{eq:Ldark} with $\cx\!=\!0$ and a representative value for $\tilde{b}_x \equiv \frac{\alpha_\textsc{x}}{8\pi}b_X$.

\section{Phenomenology of a light dilaton at the LHC}
\label{sec:pheno}
In this section, we study in details the phenomenology of a light dilaton at the LHC in a mass range $[10\!-\!300]\gev$, for the three models described in the previous section with and without the presence of the dark sector portal. 

In Tab.~\ref{tab:coeff_models}, we collect all the relevant couplings introduced in the effective Lagrangian~\eqref{eq:sint} and \eqref{eff_lagrangian} for the three benchmark models.
For convenience we define the following quantities in relation with the mixing matrix~\eqref{h-r_mixing},
\beq
g_h\!\equiv\! c_{\theta}\!+\!\bk \frac{s_{\!\theta}}{Z}, \lsp g_\phi\!\equiv\!s_{\!\theta}\!-\!\bk \frac{c_{\theta}}{Z}, \lsp \tilde g_h\!\equiv\! -\frac{v}{f} \frac{s_{\!\theta}}{Z} , \lsp \tilde g_\phi\!\equiv\! \frac{v}{f} \frac{c_{\theta}}{Z},
\label{gs_param}
\eeq
In this notation, $g_h$ and $g_\phi$ are the components of the interaction-basis SM Higgs field $h_0$ into the physical SM Higgs $h$ and the dilaton $\phi$ mass eigenbasis. 
Conversely, the $\tilde g_h$ and $\tilde g_\phi$ are the components of the dilaton in the interaction base (in Eq.~\eqref{eq:sint}) to the physical SM Higgs $h$ and the dilaton $\phi$, 
respectively, multiplied by an extra $v\!/\!f$ suppression factor. 
With this notation the Feynman rules of the model result in very compact expressions which are presented in Tab.~\ref{fig:feynrules} in Appendix~\ref{sec:feyn_rules}.
\begin{table}[t!]
\centering 
\setlength{\tabcolsep}{7pt}
\renewcommand{\arraystretch}{1.3}
\rowcolors{1}{white}{gray!7}
\begin{tabular}{|c|c|c|c|}
\hline
couplings & Minimal Dilaton&  Holographic Dilaton & Gauge-philic Dilaton  \\
\hline \hline
$\bk$ &$0$& $6\xi\frac{v}{f}$&$0$  \\ 
$\cm$ &$\cm$ &$0$  & $0$ \\ 
$\cw$& $1$ & $1-\tfrac{3\pi k R \, m_W^2}{f^2(k/\mpl)^2}$  & $0$ \\ 
$\cz$ & $1$ & $1-\tfrac{3\pi k R \, m_Z^2}{f^2(k/\mpl)^2}$  &$0$  \\ 
$\bw$ &$0$ & $\frac{2}{\alpha_\textsc{em} k R} $& $b_2\!/\!\sin^2\!\theta_\textsc{w}$\\ 
$\bz$  &$0$& $\frac{2}{\alpha_\textsc{em} k R} $& $\!(\!b_2\!/\!\tan^2\!\theta_\textsc{w}\!+\!b_1\! \tan^2\!\theta_\textsc{w}\!)\!$\\ 
$\ba$ &$(b_2\!+\!b_1)$& $(b_2\!+\!b_1)\!+\!\frac{2}{\alpha_\textsc{em} kR}$&$(b_2\!+\!b_1)$  \\ 
$\baz$&$(b_2\!/\!\tan\!\theta_\textsc{w}\!-\!b_1\! \tan\!\theta_\textsc{w})$  &$(b_2\!/\!\tan\!\theta_\textsc{w}\!-\!b_1\! \tan\!\theta_\textsc{w})$ & $(b_2\!/\!\tan\!\theta_\textsc{w}\!-\!b_1\! \tan\!\theta_\textsc{w})$\\ 
$\bg$ & $b_3$& $b_3\!+\!\frac{2}{\alpha_s kR}$& $b_3$\\ 
$\cf^i$ & $1$ &$1$ & $0$  \\ 
$\ch$ & $1$ & $(1-3\xi)$ & $0$\\ 
$\bh$ & $1$& $(1-6\xi)$& $0$ \\ 
$\cx$ & $1$ & $1$ & $0$\\ 
$\bx$ & $0$& $0$& $\bx$ \\ 
\hline
\end{tabular}
\caption{The model dependent couplings $b$'s and $c$'s of the minimal dilaton, holographic dilaton, and gauge-philic dilaton models. Here $\xi$ is the Higgs-curvature mixing parameter and $kR$ is the volume factor in the holographic model. Whereas, $\alpha_\textsc{em}$ and $\alpha_s$ are the electromagnetic  and strong  coupling constants, respectively, and $\theta_\textsc{w}$ is the Weinberg angle. Above $b_3,b_2,b_1$ are the $SU(3)_c,SU(2)_L,U(1)_Y$ gauge coupling $\beta$-function coefficients, respectively. }
\label{tab:coeff_models}
\end{table}

\begin{figure} [t!]
\centering
\includegraphics[width=0.45\textwidth]{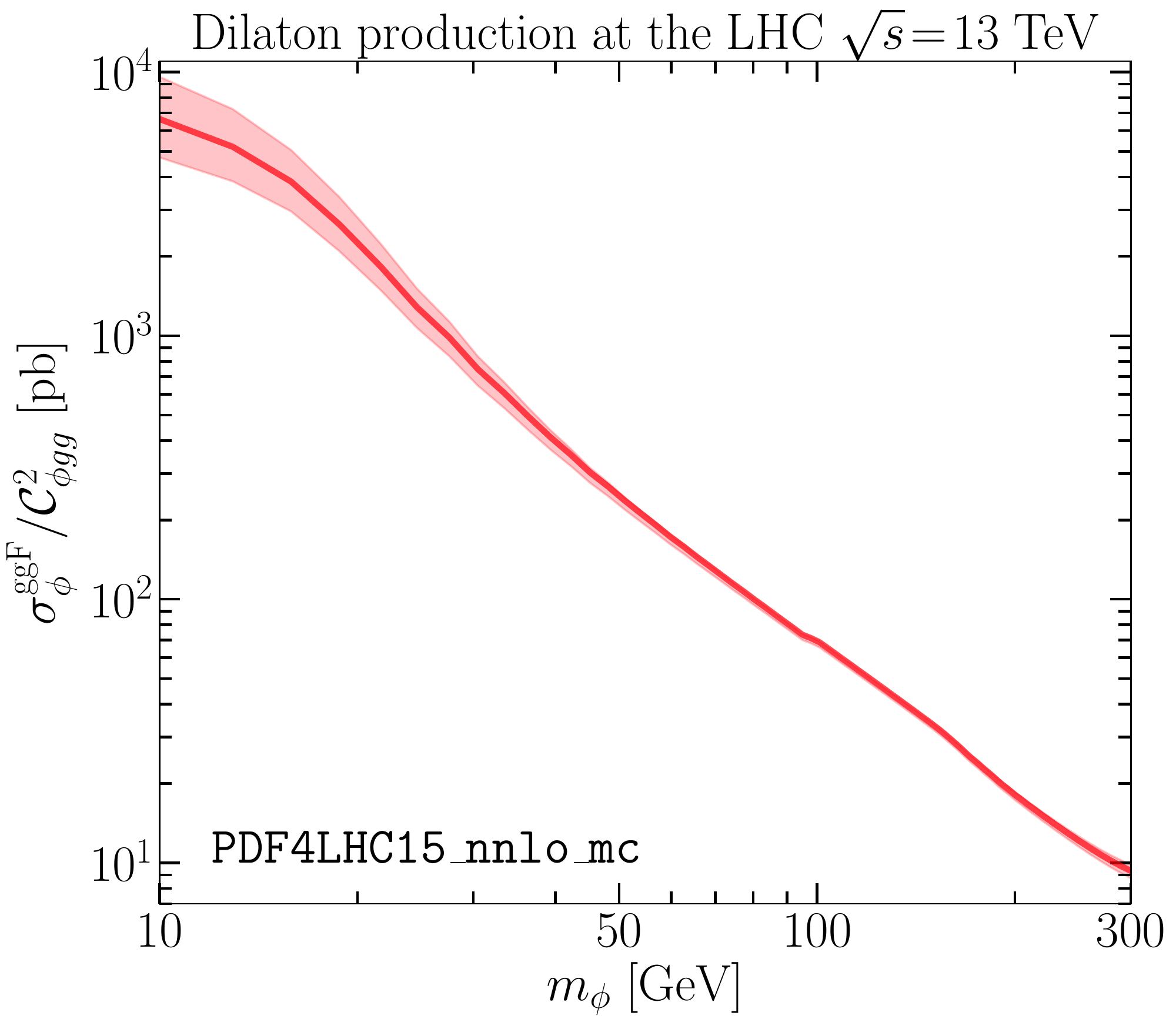}
\caption{Cross section of the dilaton $\phi$ via gluon fusion normalized w.r.t. ${\cal C}^2_{\phi gg}$ at the LHC with $13\tev$ center of mass energy. The colored bands show uncertainties associated with the \texttt{PDF}s and renormalization/factorization scale.}
\label{fig:xsecphi}
\end{figure}

\paragraph{Dilaton production cross-sections:}
The cross-sections for the dilaton can be directly obtained by employing existing information about the Higgs production cross section at different Higgs masses.
Given an initial state $i$ and a final state $j$, the resonant dilaton production cross section is given by
\beq
\sigma^{i}_{\phi\to j}\!\equiv\!\sigma(i\to\phi)\!\cdot\!{\text{BR}}(\phi\to j)=\sigma^{\text{SM}}(i\to h)\big\vert_{m_h\!=\!m_{\phi}}\!\cdot{\cal C}_{\phi i}^2  \!\cdot\!{\text{BR}}(\phi\to j).
\eeq
Above $\sigma^{\text{SM}}(i\!\to\! h)$ is the production cross-section of the SM Higgs evaluated at the dilaton mass,
${\text{BR}}(\phi\to j)$ is the dilaton branching ratio to $j$ final state, and the effective coupling ${\cal C}_{\phi i}$ is defined as
\beq
{\cal C}^2_{\phi i} \equiv \frac{\sigma(i \to\phi)}{\sigma^{\text{SM}}(i \to h)|_{m_h\!=\!m_{\phi}}}\,.
\eeq
The SM-like Higgs gluon fusion production cross section at different masses can be computed with the public code 
\texttt{SusHi}~\cite{Harlander:2012pb,Harlander:2016hcx} 
 (which 
takes into account ${\rm NNNLO}$ QCD and approximate ${\rm NNLO}$ EW corrections)
where we used \texttt{PDF4LHC15\_nnlo\_mc} parton distribution functions, and
the renormalization and factorization scales at $\mu\simeq m_\phi/2$.
In Fig.~\ref{fig:xsecphi} we plot the cross section of the dilaton $\phi$ via gluon fusion 
at the LHC with $13\tev$ center of mass energy
normalized w.r.t. ${\cal C}^2_{\phi gg}$. 
The cross sections for vector boson fusion (VBF) and single Higgs production (in association with a massive gauge boson) can be read from the Higgs Cross-section Working Group~\cite{deFlorian:2016spz}. 
We note that, as a consequence of the sizeable coupling of the dilaton to gluons through the $\beta$-function coefficient, 
the main production channel for the dilaton is gluon fusion, whereas the others are subleading in the whole parameter space considered in this work and we neglect them.
Hence the relevant coefficient to describe the dilaton production at the LHC is the 
effective dilaton coupling for gluon fusion, which is,
\beq
{\cal C}^2_{\phi gg} =\bigg\vert
\frac{2 b_g  \tilde g_\phi  -\sum_i (g_\phi+\cf^i \tilde g_\phi)F_{1\!/\!2}(\tau_i) }{\sum_{i}F_{1\!/\!2}(\tau_i) } \bigg\vert^2
\approx
\bigg\vert \frac{2b_g\tilde g_\phi}{F_{1\!/\!2}(\tau_{t})}-(g_\phi+\cf^t \tilde g_\phi) \bigg\vert^2 .\label{eq:Cphigg}
\eeq
The summation $i$ above runs over the quarks flavors in the loops and the form factor $F_{1\!/\!2}(\tau)$ is given in Appendix~\ref{sec:feyn_rules}.
The last approximation in Eq.~\eqref{eq:Cphigg} takes into account the fact that the dominant fermion loop contribution is from the top quark. 

\paragraph{Global fit to the Higgs data:}
Given that we have kinetic and mass mixing between the Higgs and the dilaton, the Higgs couplings to the other SM fields get modified.
This implies stringent constraints on dilaton models since the SM-Higgs properties (production and decay modes) have been precisely measured
at the LHC run-2.
In the following we perform global $\chi^2$ fit to the Higgs signal strengths $\mu^{i}_{j}$ in the different channels
\beq
\mu^{i}_{j}\equiv \frac{\sigma(i\to h)\!\cdot\!{\text{BR}}( h\to j)}{\sigma^{\rm SM}(i\to h)\!\cdot\!{\text{BR}}^{\rm SM}(h\to j)}={\cal C}_{h i}^2~\frac{{\text{BR}}(h\to j)}{{\text{BR}}^{\rm SM}(h\to j)}\,, 	\label{eq:signalstrength}
\eeq  
which are defined as the production times the decay rates for each initial state $i$ and final state $j$, relative to those of the SM.
The Higgs effective couplings to the $i\!=\!gg,VV,$ etc. states are encoded in the coefficients ${\cal C}_{h i}^2$.
The global $\chi^2$ fit is performed with the code \texttt{Lilith-2}~\cite{Bernon:2015hsa,Kraml:2019sis}
(see Refs.~\cite{Bernon:2015hsa,Kraml:2019sis} for details on \texttt{Lilith-2} and the experimental data used).  
The fit to Higgs data will provide a further important indirect constraint in our phenomenological analysis.

\paragraph{Details on implementation of collider bounds:}
In the following we present the current LHC constraints on the different light dilaton scenarios, based on the LHC resonance searches.
The set of LHC (and LEP) searches that we consider is listed in Tab.~\ref{tab:bounds}, where the mass range and the luminosity of a given search is mentioned.
Concerning the LEP constraints, we employ the search looking for Higgs-like state decaying into $b \bar b$ final state \cite{Barate:2003sz} and into hadrons \cite{Abdallah:2004bb}, 
which provide bounds on the ${\cal C}_{\phi ZZ}^2$ coupling times the relevant branching ratio. 
We will choose the strongest of the two constraints for every point of our parameter space and we will show only one common line for the LEP bounds.
\begin{table}[t!]
\centering
\begin{tabular}{|c|l|c|c|c|}
\hline
Experiment           &Decay channels & Mass range [GeV] & Luminosity[fb$^{-1}$]& Reference    \\
\hline\hline
LEP&        ${\cal C}_{\phi Z Z}^2$          &      $1-115$   &$[2.46,0.61]$& \cite{Barate:2003sz,Abdallah:2004bb}   \\
             
 \hline 
ATLAS &        $\gamma\gamma $          &      $65-600$   & 20.3~(8\tev)& \cite{Aad:2014ioa}   \\
                       &    $\gamma\gamma $    &      $ 65-110$    &  80& \cite{ATLAS:2018xad}   \\
           &        $hh$         &  $260-500$        &36.1&   \cite{Aaboud:2018ewm}  \\
           &        $Z\gamma$       &    $250-2400$       &36.1 &   \cite{Aaboud:2017uhw}   \\
\hline
CMS      & $jj$          &    $50-300$      & 35.9 &   \cite{Sirunyan:2017nvi}   \\
                       &    $\gamma\gamma $    &      $ 70 -110$      &19.7~(8\tev), 35.9& \cite{Sirunyan:2018aui}   \\    
                       &    $\gamma\gamma $    &      $ 150-850 $   &19.7~(8\tev)  & \cite{ Khachatryan:2015qba}   \\
          &        $hh$         &  $260-500$         & 35.9&   \cite{Sirunyan:2018two}  \\
           &   $ZZ\rightarrow llll$           &    $140-2500$    & 12.9    &  \cite{CMS:2016ilx}    \\
           &       $WW$          &     $200-3000$      & 35.9&  \cite{CMS:2019kjn}    \\
\hline
\end{tabular}
\caption{This table collects all the relevant analyses used in our phenomenological study of a light dilaton with their respective mass range and luminosity, including LEP, ATLAS and CMS results. The ATLAS and CMS results are mainly from the LHC run-2 with the center of mass energy $13\tev$, whereas the ones from $8\tev$ are indicated.}
\label{tab:bounds}
\end{table}

At the LHC there are several searches for a Higgs-like state decaying into different channels, that we employ in our analysis\,\footnote{We neglect possible effects due to interference between the SM-Higgs and the dilaton production, that could occur when $m_{\phi}\simeq 125$ GeV.}.
When available, we used the reported limits corresponding to the gluon fusion production.
For final states covered both by ATLAS and CMS searches, in our plots we will display the resulting stronger limit.
For the low mass di-jet resonant search of CMS \cite{Sirunyan:2017nvi}, where the original interpretation is based on a model with quark-antiquark production, 
we used a conversion factor as derived in \cite{Mariotti:2017vtv} to convert to a limit for a gluon fusion process. In order to asses the mono-jet reach of the analysis \cite{Aaboud:2017phn}, we have implemented our model in \texttt{FeynRules}~\cite{Alloul:2013bka} and computed the efficiency of emitting one hard extra jet with \texttt{MadGraph5}~\cite{Alwall:2014hca,Alwall:2011uj}.
We consider all the signal categories of \cite{Aaboud:2017phn} and we select the strongest in order to draw the sensitivity lines. We argue that this simplified procedure is sufficient to identify the reach of the LHC monojet in our phenomenological analysis.
When relevant, we also add the bound obtained in the phenomenological analysis of \cite{Mariotti:2017vtv}, based on the experimental public data from di-photon cross section measurements.
We label this bound as $\gamma \gamma_{\textsc{mrst}}$.
Finally, the high-luminosity LHC (HL-LHC) sensitivity curves are derived by taking the current expected sensitivities at 95\% CL from the experimental papers 
and by rescaling them with the square root of the luminosity
(counting a total of $3000 \text{fb}^{-1}$), 
that is by assuming that the dominant uncertainty on the SM background is statistical.

\paragraph{Remarks on the different dilaton mass regions:}
In the following we study the LHC phenomenology of the three different benchmark dilaton models explained in the previous section, with or without the dark sector portal,
focusing on the less explored
dilaton mass range~$[10\!-\!300]\gev$.
Within this mass window, in order to characterize the relevant LHC signatures, it is instructive to separate three mass ranges: 
\beq 
R_1=[10\!\sim\!60]\gev , \lsp R_2=[60\!\sim\!160]\gev, \lsp R_3=[160\!\sim\!300]\gev.		\label{eq:R123}
\eeq
The range $R_1$ covers masses below $m_h/2$. Here LHC has currently limited searches and as a result the LEP bounds become very important. In this mass range more dedicated experimental analysis could improve the LHC reach and hence could constitute a promising near future discovery channel for a light dilaton.
The mass range $R_2$ is between $m_h/2$ and $2m_W$, and it is relatively much better covered at the LHC. In particular in this range di-photon searches typically put stringent limits on the interaction scale $f$. 
Finally, for dilaton masses above $2m_W$, i.e. the region $R_3$, the dilaton couples and can decay into massive gauge bosons analogously as a would-be-heavy Higgs boson, and therefore these decay channels provide very strong constraints on scales $f$ up to $\op(5\!-\!10)\tev$.

\subsection{Minimal dilaton}
In this scenario, we assume that the dilaton has no kinetic mixing with the Higgs boson. However, we include a mass mixing of the form $\cm m_{\phi_0}^2 \phi_0 h_0$ in the effective Lagrangian~\eqref{eff_lagrangian}. 
In this case the mixing matrix is given by~\eqref{eq:massmixing}, where the mixing angle is parameterized in terms of $\cm$ and the dilaton mass. 
For concreteness, in the following we consider two cases with $\cm\!=\!0$ (no mass mixing) and $\cm\!=\!0.1$. 
In the case $\cm\!=\!0.1$ the mixing angle asymptotes to $\sin^2\theta\to 0$ and $\sin^2\theta\to 1\%$ in the limits $\mphi^2\!\ll\!\mh^2$ and $\mphi^2\!\gg\! m_h^2$, respectively, see also Eq.~\eqref{eq:thetamm}, while the mixing angle increases as the dilaton and Higgs masses become degenerate. 
Regarding the dilaton couplings to the massless gauge bosons, we assume the values of the gauge $\beta$-function coefficients as those of the SM fields, i.e. $b_3\!=\!7$, $b_2\!=\!19/6$, and $b_1\!=\!-41/6$.
\begin{figure} [t!]
\centering
\includegraphics[width=0.5\textwidth]{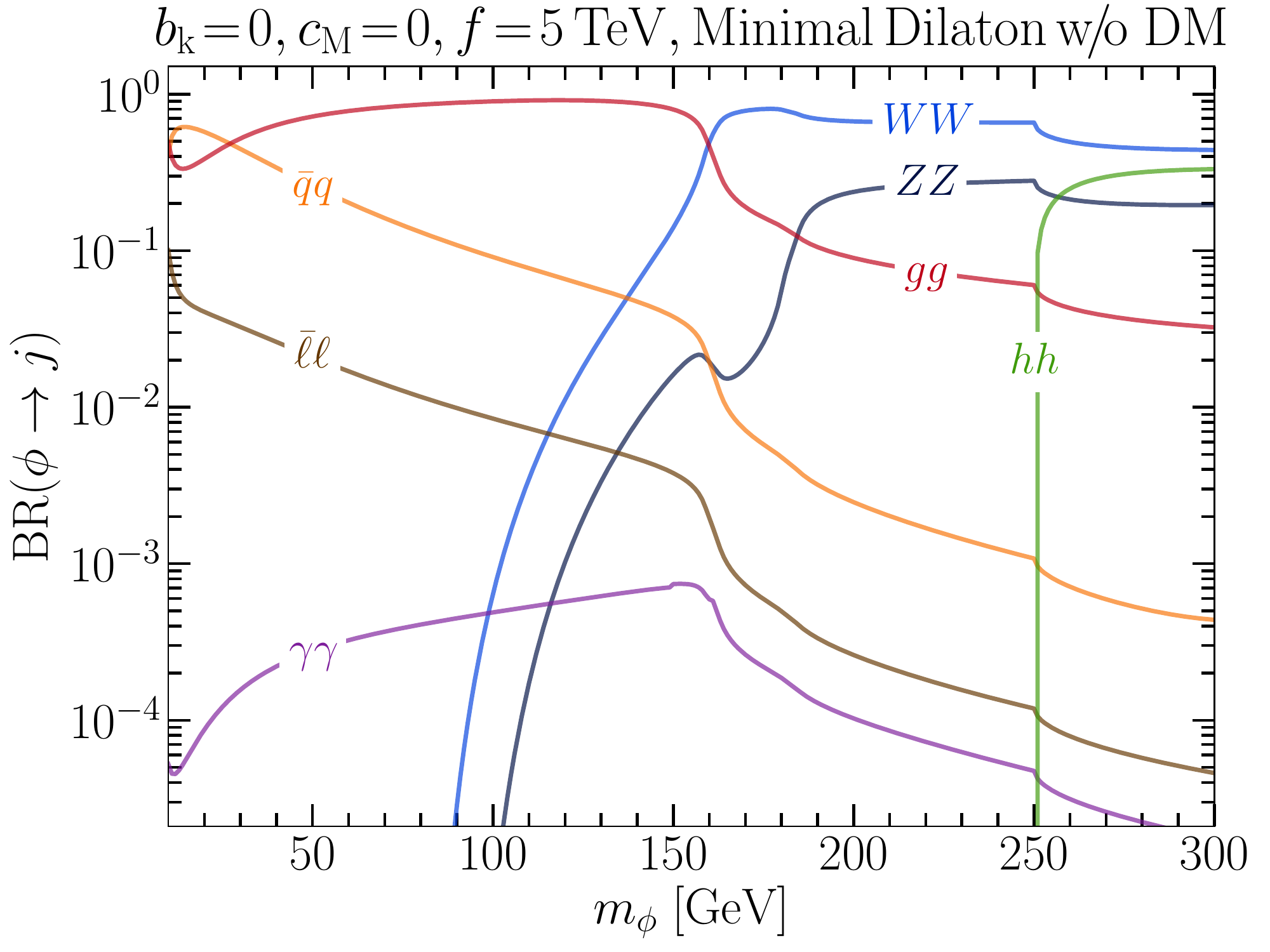}\hspace{-5pt}
\includegraphics[width=0.5\textwidth]{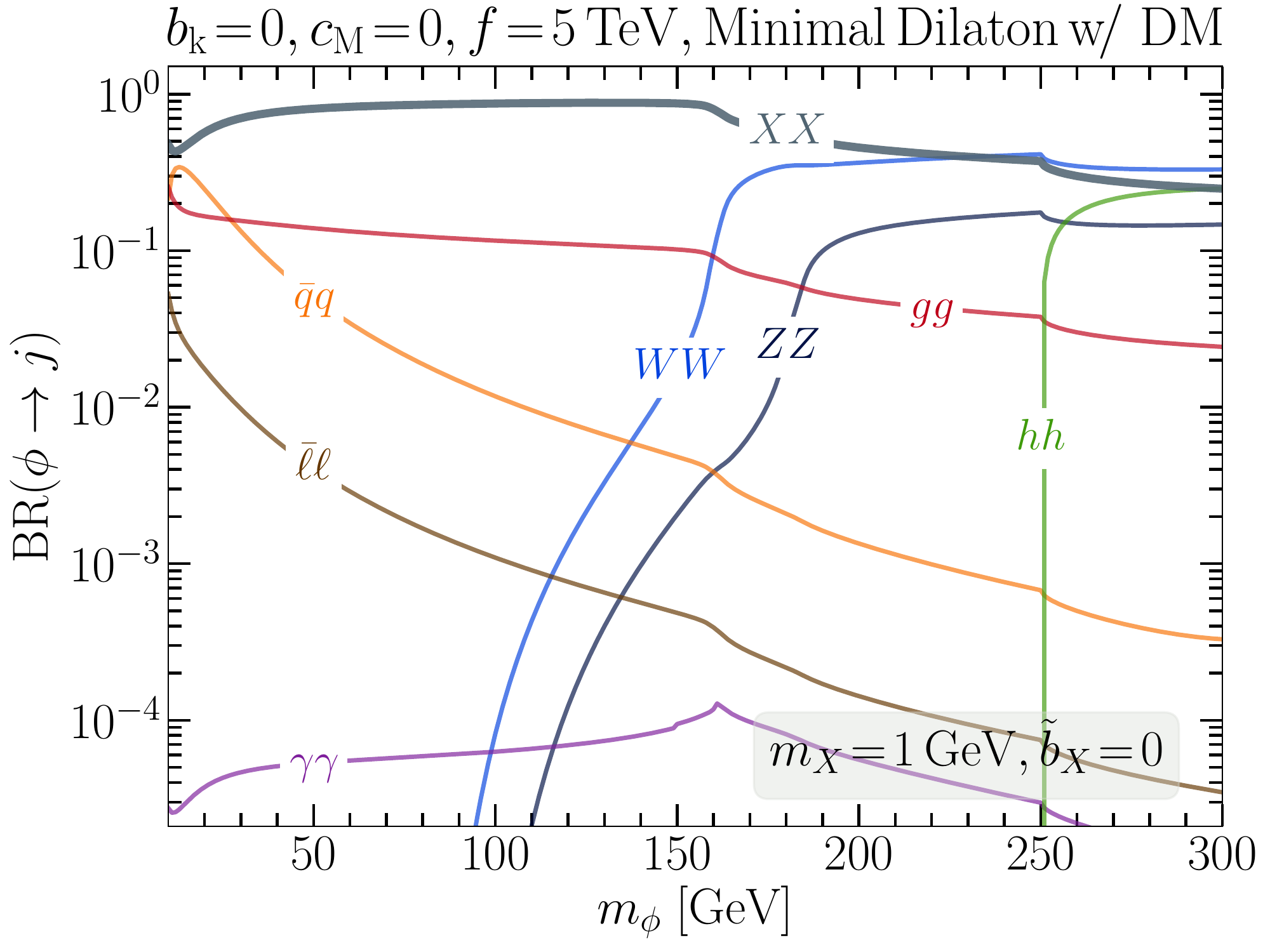} \\
\includegraphics[width=0.5\textwidth]{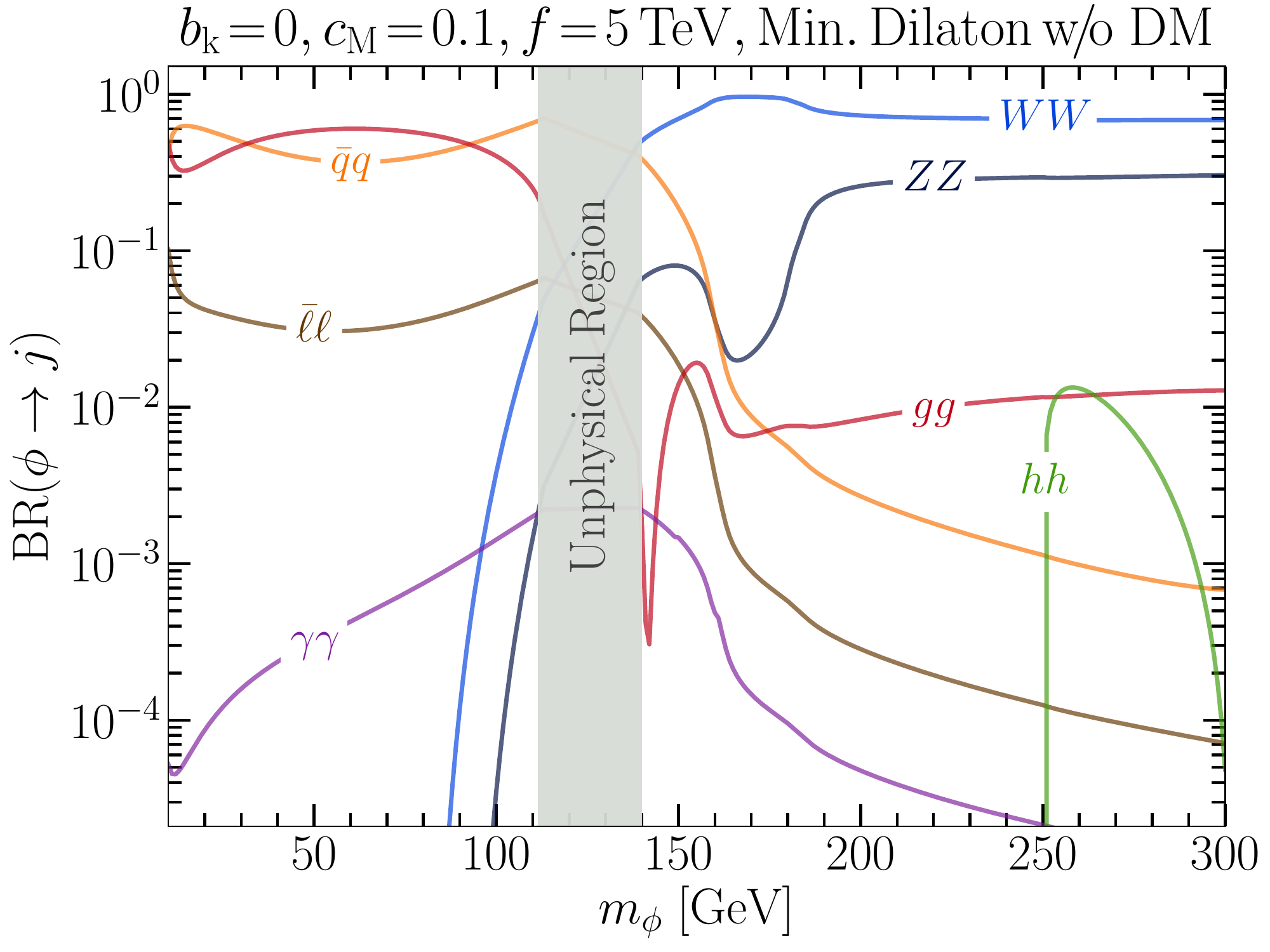}\hspace{-5pt}
\includegraphics[width=0.5\textwidth]{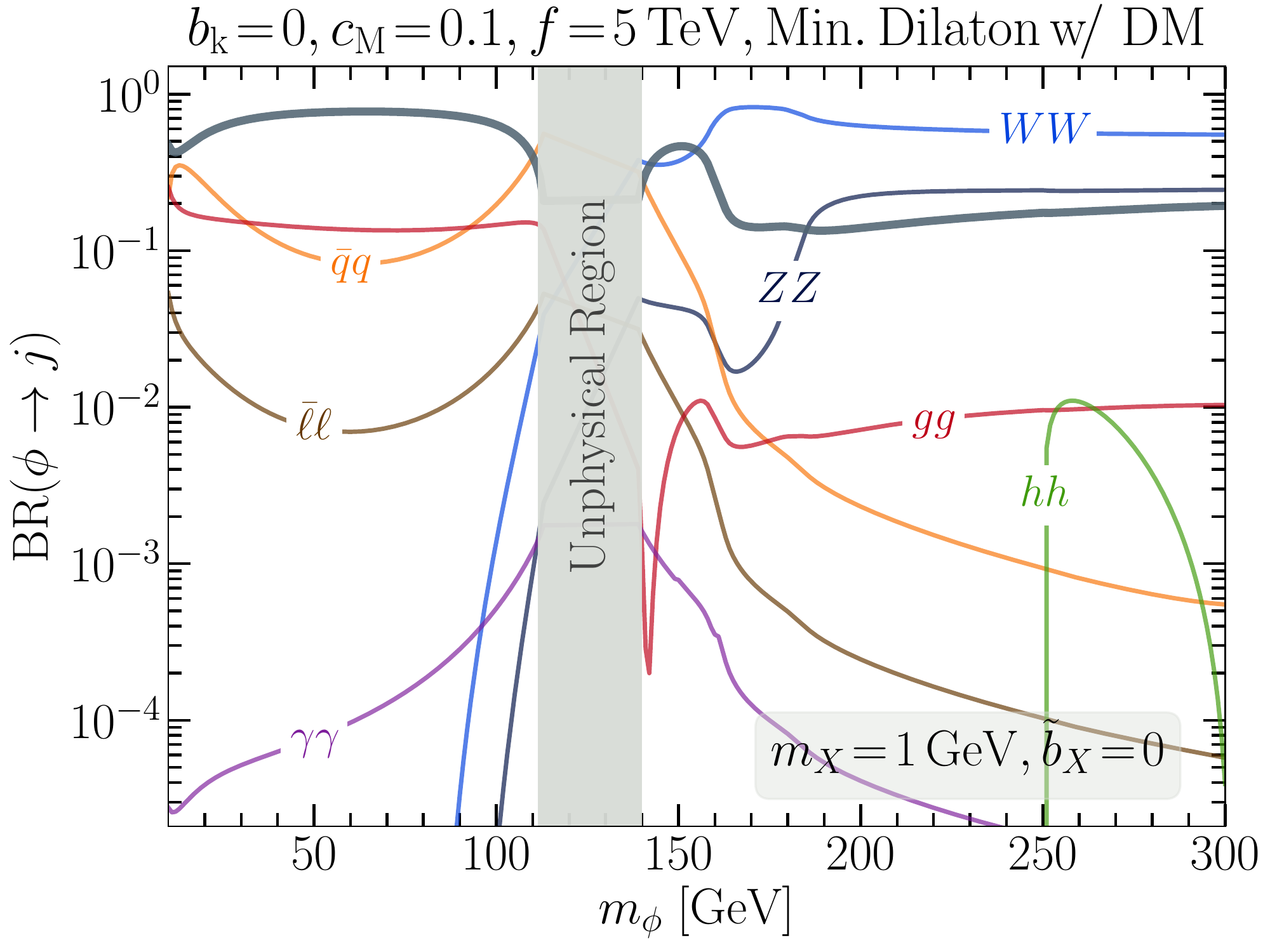} 
\caption{Branching fractions of the dilaton in the minimal dilaton scenario as a function of dilaton mass $m_\phi$ for different choices regarding the dark sector portal. The upper (lower) panel corresponds to the value of mass mixing parameter $\cm\!=\!0\, (0.1)$. The gray band in the lower-panel plots corresponds to the unphysical region where the correct Higgs and dilaton masses cannot be reproduced for the given dilaton-Higgs mixing.}
\label{fig:mm_brs}
\end{figure}

\paragraph{Branching ratios:}
In Fig.~\ref{fig:mm_brs}, the upper and lower panels show the branching ratios of the dilaton to the various final states for the mass mixing parameters $\cm\!=\!0$ and $\cm\!=\!0.1$, respectively. 
The left- (right-) panels show the branching ratios of the dilaton to the various final states in the absence (presence) of the dilaton-dark sector portal interaction. 
The grey band shows unphysical region where the masses of the SM-like Higgs and dilaton can not be reproduced from the model parameters with mixing $\cm\!=\!0.1$. 
In these plots we also fixed the scale of conformal breaking $f=5\tev$, but the dependence of the branching ratios on $f$ is very mild.
For the dilaton decay to massive vector bosons, we also include the off-shell 3- and 4-body decays since they are relevant to derive the collider bounds (in particular the $ZZ\to \ell\ell\ell\ell$ off shell decay).

The upper-left panel of Fig.~\ref{fig:mm_brs} corresponds to no dilaton-Higgs mixing scenario and without the presence of dark portal couplings. 
In this case, the dominant branching fraction of the dilaton is to gluons for $m_\phi\lesssim 2m_W$, i.e. in the mass ranges $R_1$ and $R_2$ (see~\eqref{eq:R123}). 
However, for the dilaton mass above $2m_W$, i.e. the $R_3$ region, the dominant branching fractions are to the SM massive bosons when kinematically allowed.  
The branching fraction of all the light quark flavors are summed and represented with $\bar qq$ curve.
Since the dilaton couplings are proportional to the fermion masses, the largest contribution is into bottom quarks.
Similarly, we show the leptonic branching ratio $\bar \ell \ell$ by summing over all leptons, where the largest contribution is into $\tau \tau$. 
Note that the di-photon branching fractions are considerably larger than a SM-like Higgs at the dilaton mass. 
This is mainly due to the fact that the dilaton-photon coupling receives additional contributions from the running of gauge coupling proportional to $b_{\gamma}$. 

The upper-right panel of Fig.~\ref{fig:mm_brs} corresponds to no dilaton-Higgs mixing but with the presence of a portal to the dark sector.
As mentioned, here we consider vector DM mass $m_X\!=\!1\gev$ and the other dark sector parameters are set to $\cx\!=\!1$ and $\bx\!=\!0$. 
In this case, the dominant decay mode for dilaton masses in the region $R_1$ and $R_2$, i.e. $m_\phi\!\lesssim\! 2m_W$, is into dark photon.
For high mass region $m_\phi\!>\! 2m_W$, instead, the invisible branching fraction gets comparable to the one into massive bosons $W^\pm,Z, h$. 
Note that in the low mass regions ($R_1$ and $R_2$) the dilaton branching fraction into gluons is $\op(10)\%$ and it drops to $\op(5)\%$ in the high mass region $R_3$.

In the lower panels of Fig.~\ref{fig:mm_brs} we show the branching ratios in the case of dilaton-Higgs mixing set by $\cm =0.1$ and without (left)
or with (right) the presence of a portal to the dark sector.
The dominant branching ratios in the low and high mass regions are similar to the case without the mixing. 
In particular, in the case with dark sector coupling, the branching fraction to the dark photons is dominant in the low mass region, while in the high mass region the SM massive gauge bosons branching ratios are the largest. 

Note that in the region where $|m_\phi-m_h|\lesssim25\gev$ the mixing angle becomes large and there are non-trivial cancellations in the effective couplings (precisely between the Higgs mixing contribution $g_\phi$ with the pure dilaton contribution $\tilde g_\phi$), leading to sudden drops of some of the decay modes.
In particular, we note that the branching ratio into gluons (very relevant for LHC phenomenology) has a sharp dip at around $\mphi\!\sim\!141\gev$~\footnote{Similarly, for instance the dilaton to di-Higgs branching ratio has a sharp dip at around $\mphi\!\sim\!295\gev$ for $f\!=\!5\tev$.}.
We can analytically understand the accidental cancelation in the gluon channel by inspecting the effective dilaton-gluons coupling ${\cal C}_{\phi gg}$~\eqref{eq:Cphigg}. 
In this scenario, we have $b_g\!=\!b_3\!=\!7$ and the top quark loop function can be approximated as $F_{1/2}(\tau_t)\simeq-4/3$. Hence the effective dilaton-gluons coupling is
\beq
{\cal C}^2_{\phi gg} \simeq \Big\vert g_\phi+ \tfrac{23}{2}\tilde g_\phi\Big\vert^2=\Big\vert \sin\theta+ \tfrac{23}{2}\tfrac{v}{f}\cos\theta\Big\vert^2.	\label{eq:Cphigg_mm}
\eeq
In Fig.~\ref{fig:mm_brs} we fix $f\!=\!5\tev$, therefore, ${\cal C}_{\phi gg}$ vanishes around $\mphi\!\sim\!141\gev$ where the mixing angle $\sin\theta$ is negative for $\cm\!=\!0.1$, see Eq.~\eqref{eq:thetamm}. 
Note that in such region the dilaton-Higgs mixing angle becomes quite large, and hence this region will be severely constrained by Higgs coupling measurements.
\begin{figure} [t]
\centering
\includegraphics[width=0.5\textwidth]{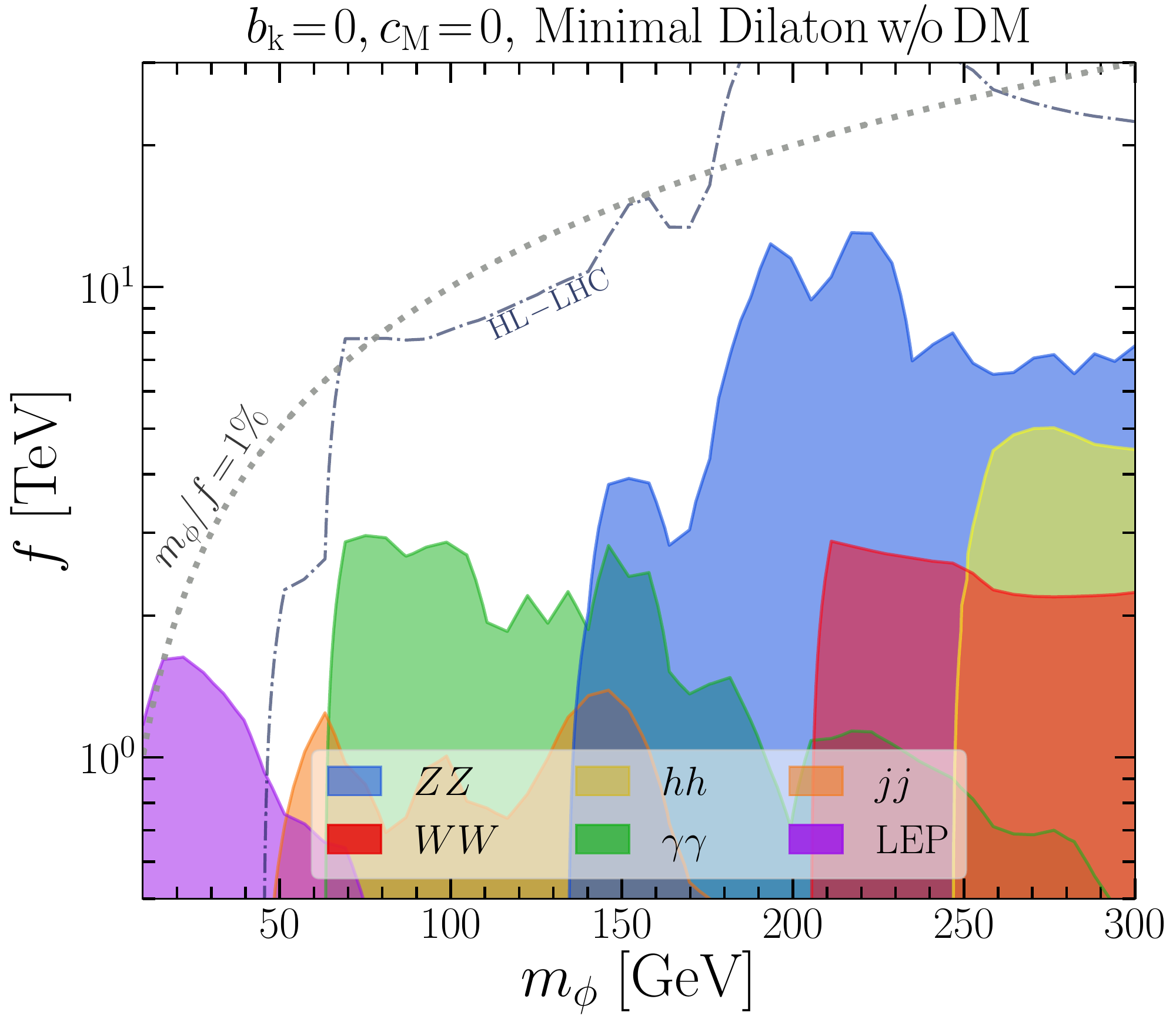}\hspace{-5pt}
\includegraphics[width=0.5\textwidth]{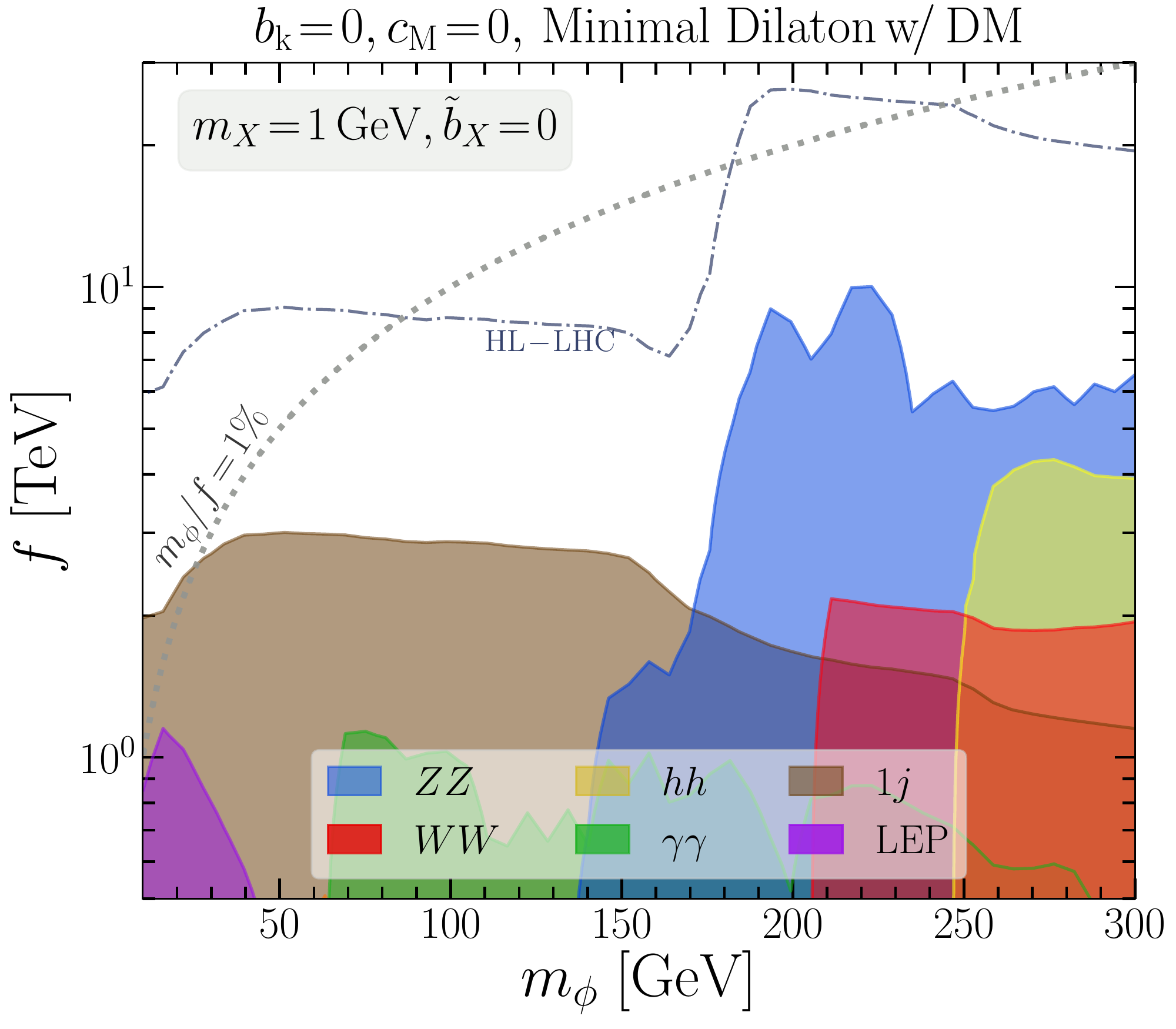}\\
\includegraphics[width=0.5\textwidth]{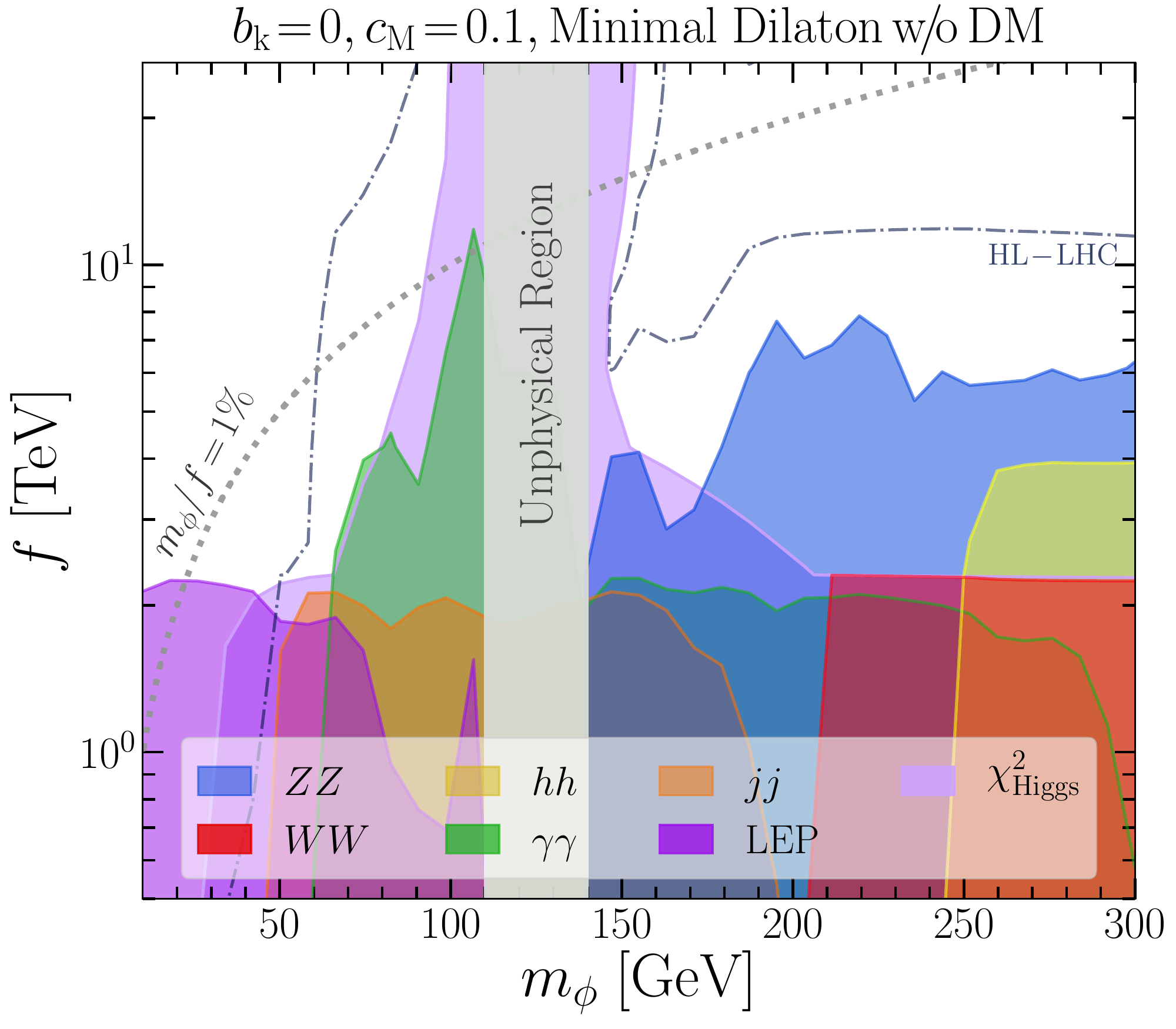}\hspace{-5pt}
\includegraphics[width=0.5\textwidth]{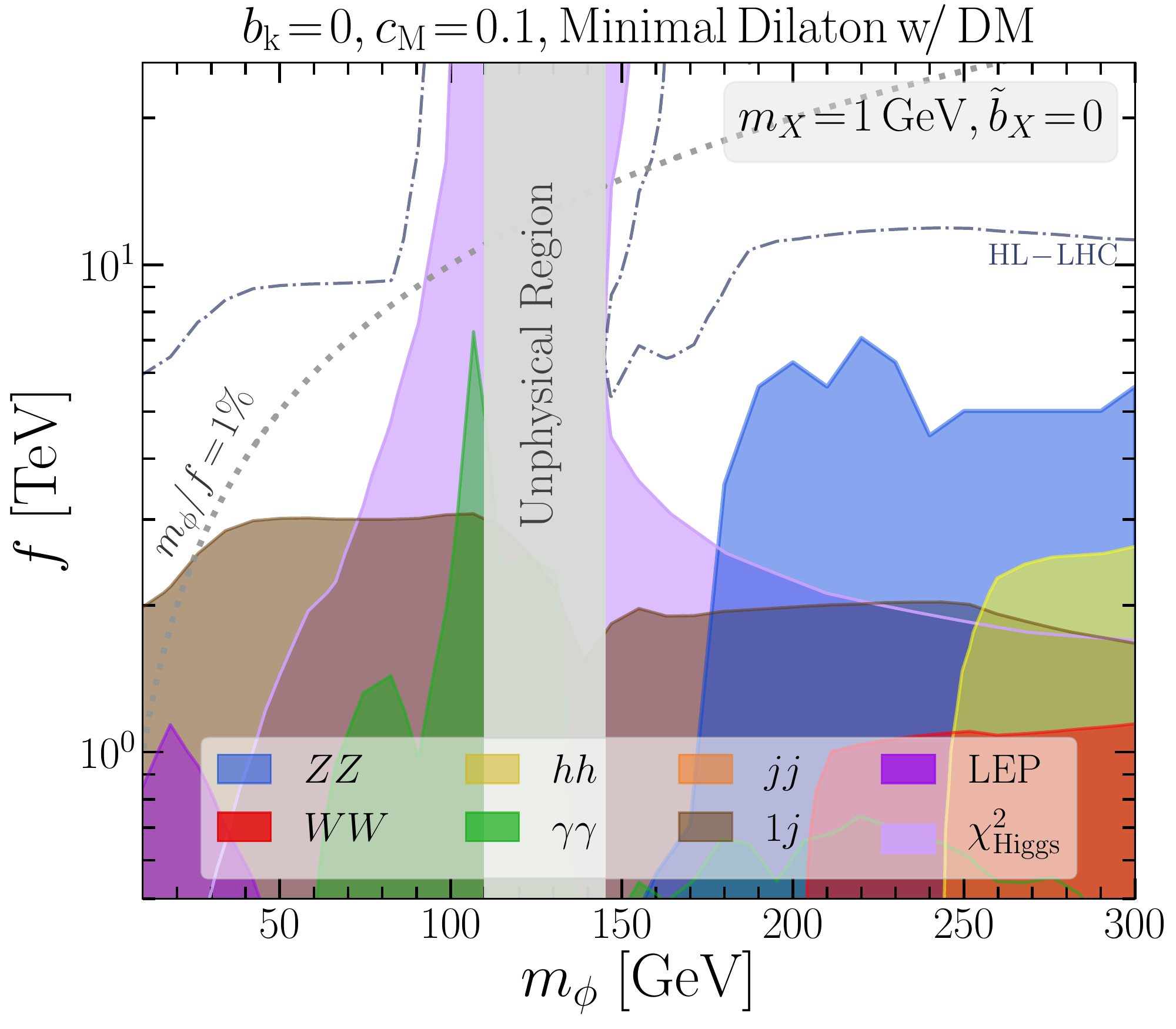}
\caption{The upper (lower) plots show the exclusion bound on the minimal dilaton parameter space in the $m_\phi$ vs $f$ plane with the mass mixing parameter $\cm\!=\!0\, (0.1)$. The color coded legends correspond to different exclusions bounds from the LHC and LEP experiments, see text. The left-panel and right-panel shows the constraints without and with the portal to the dark sector, respectively. The gray band in the lower-panel plots corresponds to the unphysical region.}
\label{fig:mmdil}
\end{figure}
\paragraph{Collider constraints:}
We present now the collider reach on the parameter space of the minimal dilaton model.
Analogously to what we did for the branching ratio
plots,
we fix the couplings as in Tab.~\ref{tab:coeff_models} and we consider the cases with $\cm \!=\!0$ and $\cm \!=\!0.1$, with and without the dark sector portal.

The remaining free parameters are the scale of conformal breaking $f$ (which determines the strength of the dilaton interactions with the SM fields) and the dilaton mass.
We thus present our collider analysis in the dilaton mass $m_{\phi}$ vs $f$ plane
as it can be seen in Figure \ref{fig:mmdil}.
The shaded colored areas are excluded because of LEP, Higgs coupling fit, ATLAS or CMS searches 
with current available public data, as it is explained in the legend of the plots. 

We begin our discussion with the case of vanishing mixing and no dark sector portal, corresponding to the left upper plot in Fig \ref{fig:mmdil}.
It is convenient to discuss these limits in the three dilaton mass regions $R_{1,2,3}$ defined in~\eqref{eq:R123}. 
We can see that the mass range $R_1\!\in\![10\!\sim\!60]\gev$ is less constrained and the most stringent direct search bound comes from the LEP experiments.
This is the region where the LHC could, with dedicated analysis, potentially improve the searches either in the di-photon or in di-jet final states. 
In the dilaton mass range $R_2\!\in\![60\!\sim\!160]\gev$, the LHC direct searches in the di-photon and di-jet channels\,\footnote{We checked that the $\mu \mu$ and $\tau \tau$ channels do not provide any additional LHC constraints.} are dominant and set a bound around $f\!\gtrsim\! 3\tev$. 
Note that the di-photon line includes both $8$ and $13$ TeV searches. In particular, the region where $m_{\phi} \!\sim\! 125$ GeV is covered only by an $8$ TeV analysis (ATLAS) and this explains the small dip in sensitivity in that mass region.
Finally, in the mass region $R_3\!\in\![160\!\sim\!300]\gev$ the strongest limits are from the di-boson ($WW,ZZ,hh$) searches, constraining $f\!\sim\!10\tev$. 
The dash-dotted curve corresponds to the reach of HL-LHC direct searches with $3000\,{\rm fb}^{-1}$ luminosity at 95\% CL, where we consider the envelope enclosing all the channels. It shows that the scale $f$ can be probed about three times stronger.
We note that at present there is a sizeable portion of the parameter space where $m_\phi\!/\!f\!>\!1\%$ which is still allowed by experiments. This will be however almost completely
covered by HL-LHC.

In the upper-right panel of Fig.~\ref{fig:mmdil} we then consider the case of no dilaton-Higgs mixing but in the presence of the dark sector portal. 
The dilaton portal to the dark sector has two important phenomenological consequences: (a) it reduces the branching fractions of the dilaton to the visible sector which leads to the weakening of the
bounds derived from SM decay products, and (b) it makes the mono-jet searches at the LHC relevant for the dilaton phenomenology. 
As we have observed, in the mass regions $R_{1,2}$ the dominant branching ratio is into the dark photons.
This implies that the most stringent bounds in these mass regions are now consequence of the mono-jet signature, leading again to a lower bound on the scale $f$ of around $3\tev$.
In the mass region $R_3$ the SM di-bosons bounds are still the strongest when kinematically allowed and set $f\!\gtrsim\!8\tev$. 
Note however that the presence of the dark sector portal weakens slightly the constraints from visible channels. 
Also in this scenario, the region where $m_\phi\!/\!f\!>\!1\%$ will be completely probed at HL-LHC.

In the lowest plots of Fig.~\ref{fig:mmdil} we then display the constraint in the case of dilaton-Higgs mass mixing.
The relative relevance of the various direct searches in the dilaton mass regions are analogous to the unmixed case.
The mass mixing implies however that the region where $|m_\phi-m_h|\lesssim25\gev$ is severely constrained by the Higgs data (or it is unphysical) because of the
large dilaton-Higgs effective mixing (see Eq.~\eqref{eq:massmixing}).
Note that the constraints deriving from Higgs measurements are not symmetric around the unphysical band. This is due to the fact on our benchmark (with fixed $\cm$) the mixing angle
scales differently for $m_{\phi}$ larger or smaller than $m_{h}$, as discussed in equation \eqref{eq:thetamm}.

Once again, the HL-LHC reach can cover essentially all the portion of parameter space with $m_\phi\!/\!f\!>\!1\%$.
As a last remark, we note that for $\cm\!>\!0.1$ the direct and indirect constraints get stronger and when $\cm\!\gtrsim\!0.3$ they cover all the interesting parameter space already with current LHC data. 

\subsection{Holographic dilaton}
In the holographic dilaton scenario we have four free parameter $m_\phi, f, \xi$ and $\mkk[f,k/\mpl]$. 
The natural value of the non-minimal dilaton-Higgs mixing parameter $\xi$ is $\op(1)$, see e.g.~\cite{Giudice:2000av}. 
In this work we choose two cases: $\xi\!=\!0$ (no kinetic mixing) and  $\xi\!=\!1/6$ (conformal point). 
The first one will serve as a reference point to compare with the other scenarios without dilaton-Higgs mixing.
As explained in Section~\ref{sec:holodil} the KK mass scale $\mkk$ is a function of $f$ and $k/\mpl$.
In our phenomenological analysis we choose the parameter $k/\mpl$ such that the scale of $\mkk$ is set to $4\tev$, making 
the KK resonances beyond the LHC reach.
We discuss our results as a function of the remaining two free parameters $m_\phi\in[10\!-\!300]\gev$ and $f$. 
As mentioned, the parameter $k/\mpl$ can not be taken arbitrarily large for consistency of the 5D action and this sets indirectly (since we fixed $\mkk=4\tev$) a lower bound on $f\gtrsim 1.6 \tev$ that we will display in our summary plots.

\begin{figure} [t!]
\centering
\includegraphics[width=0.5\textwidth]{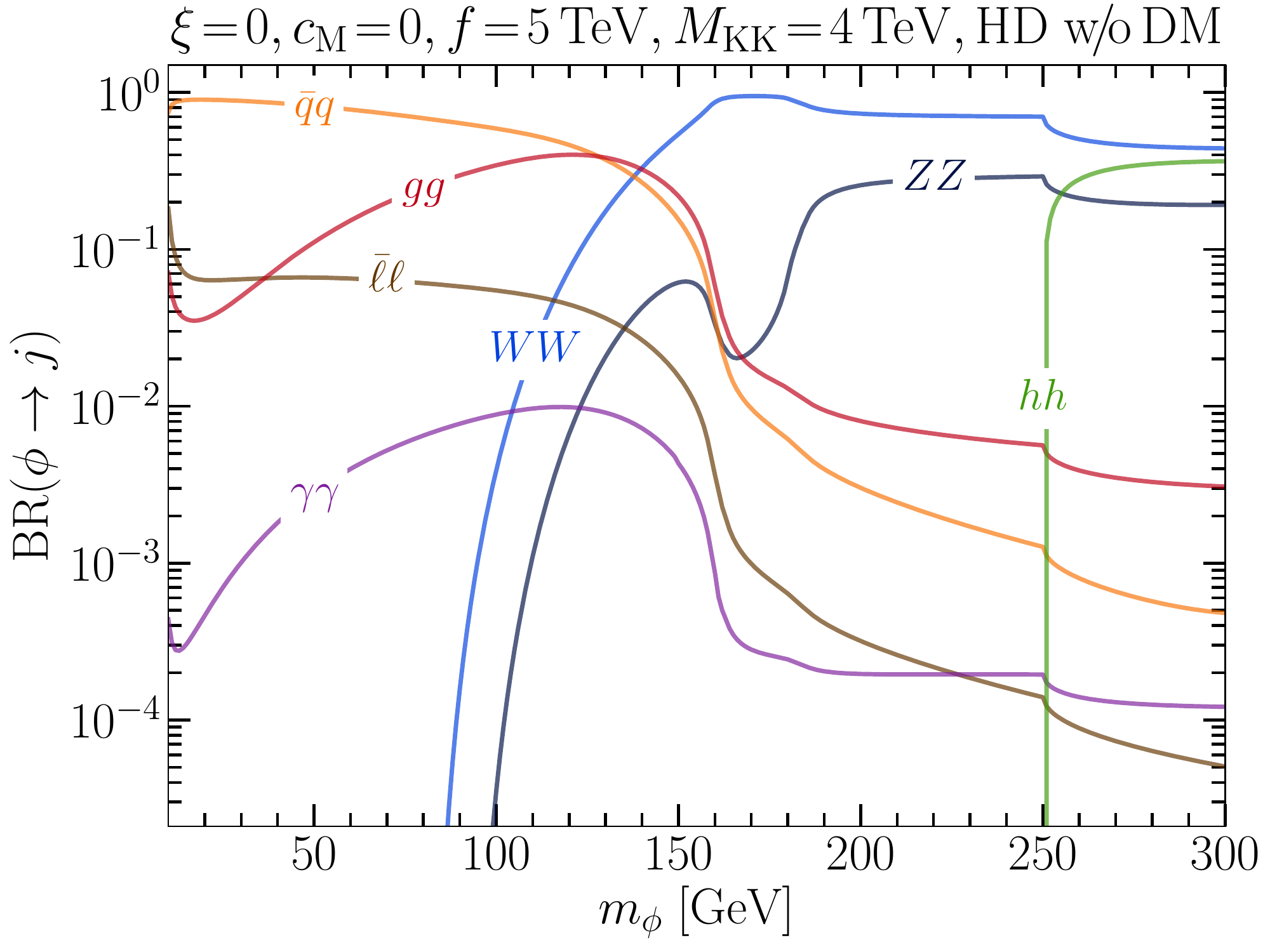}\hspace{-5pt}
\includegraphics[width=0.5\textwidth]{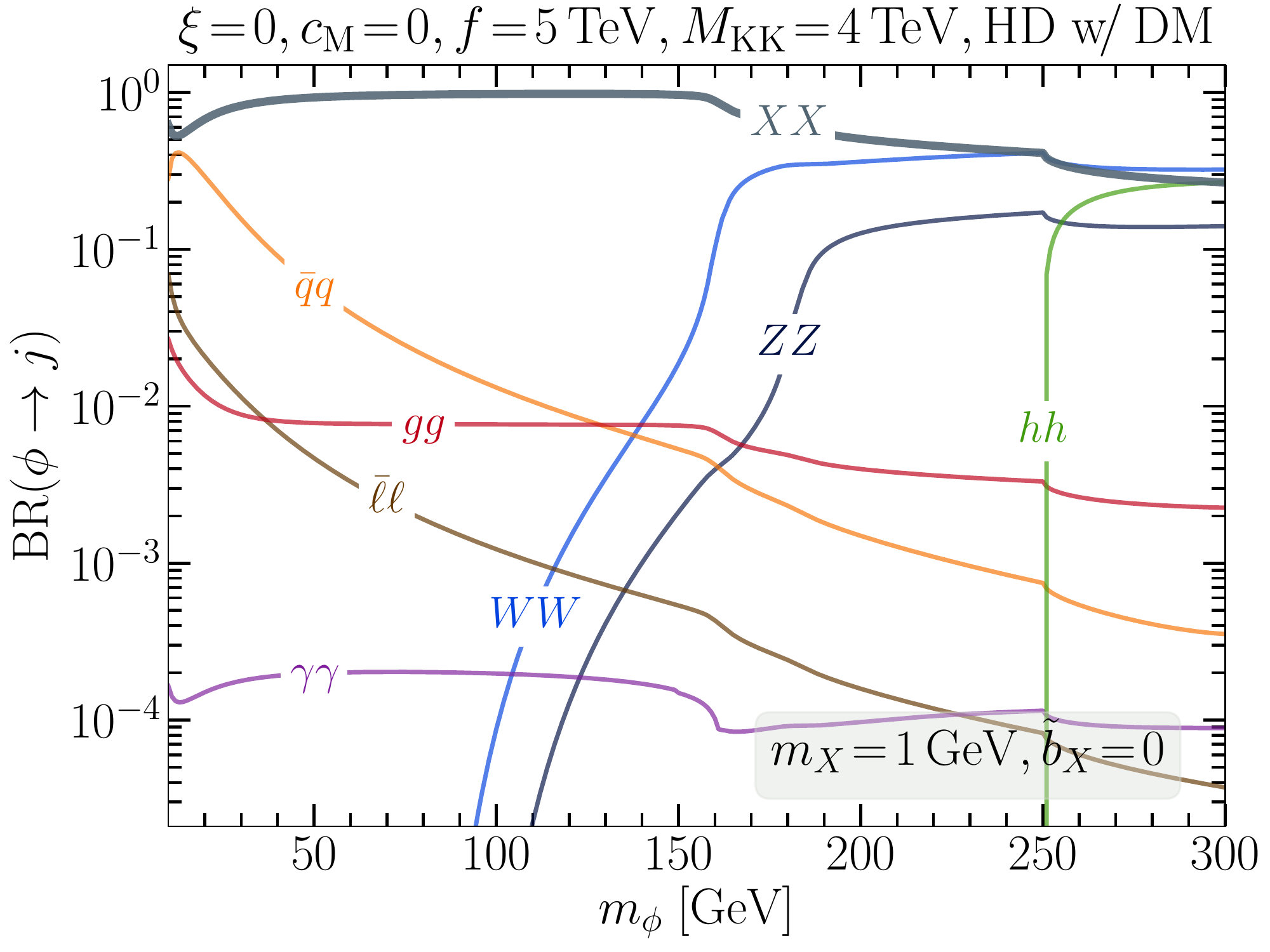}\\
\includegraphics[width=0.5\textwidth]{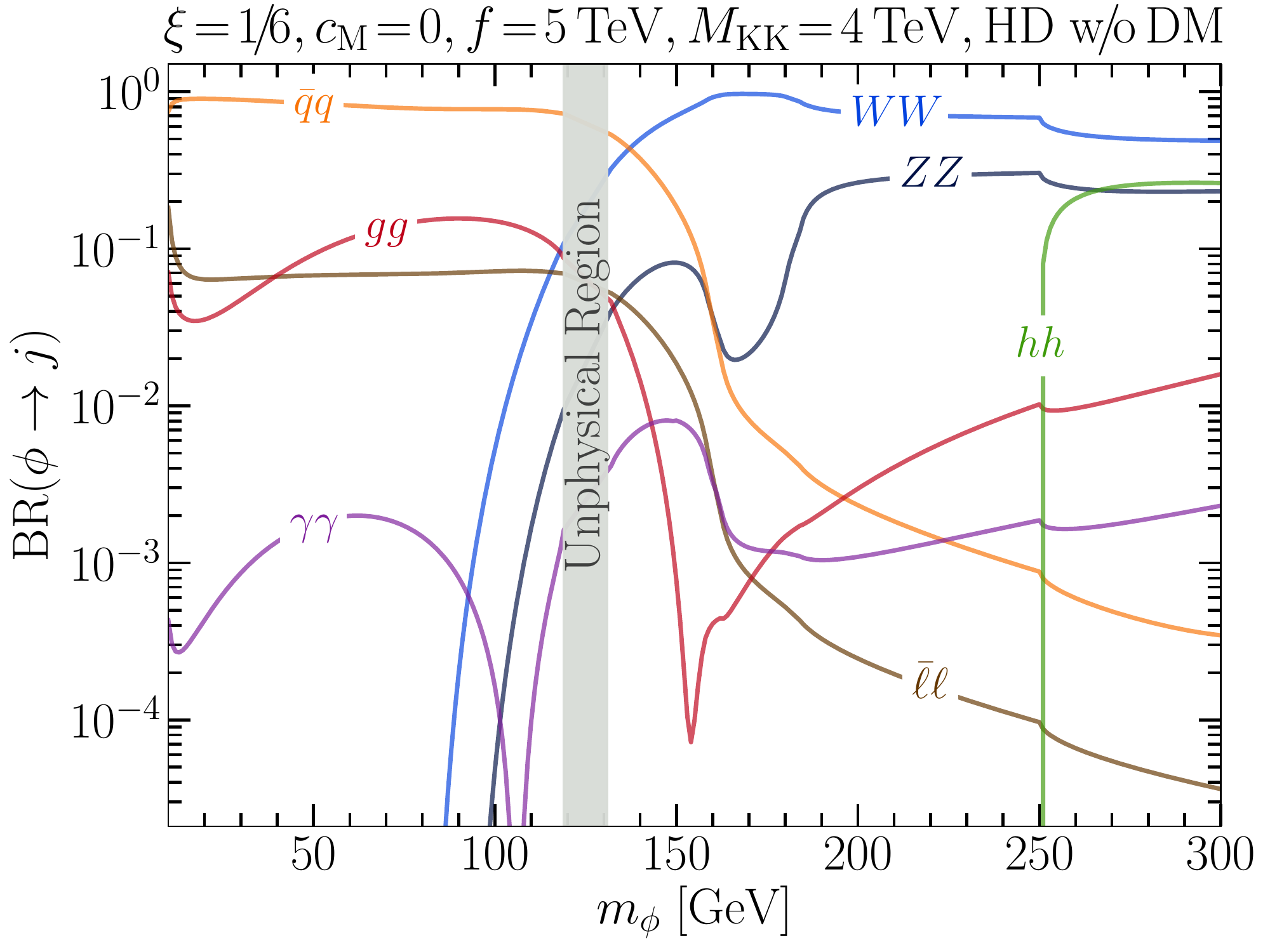}\hspace{-5pt}
\includegraphics[width=0.5\textwidth]{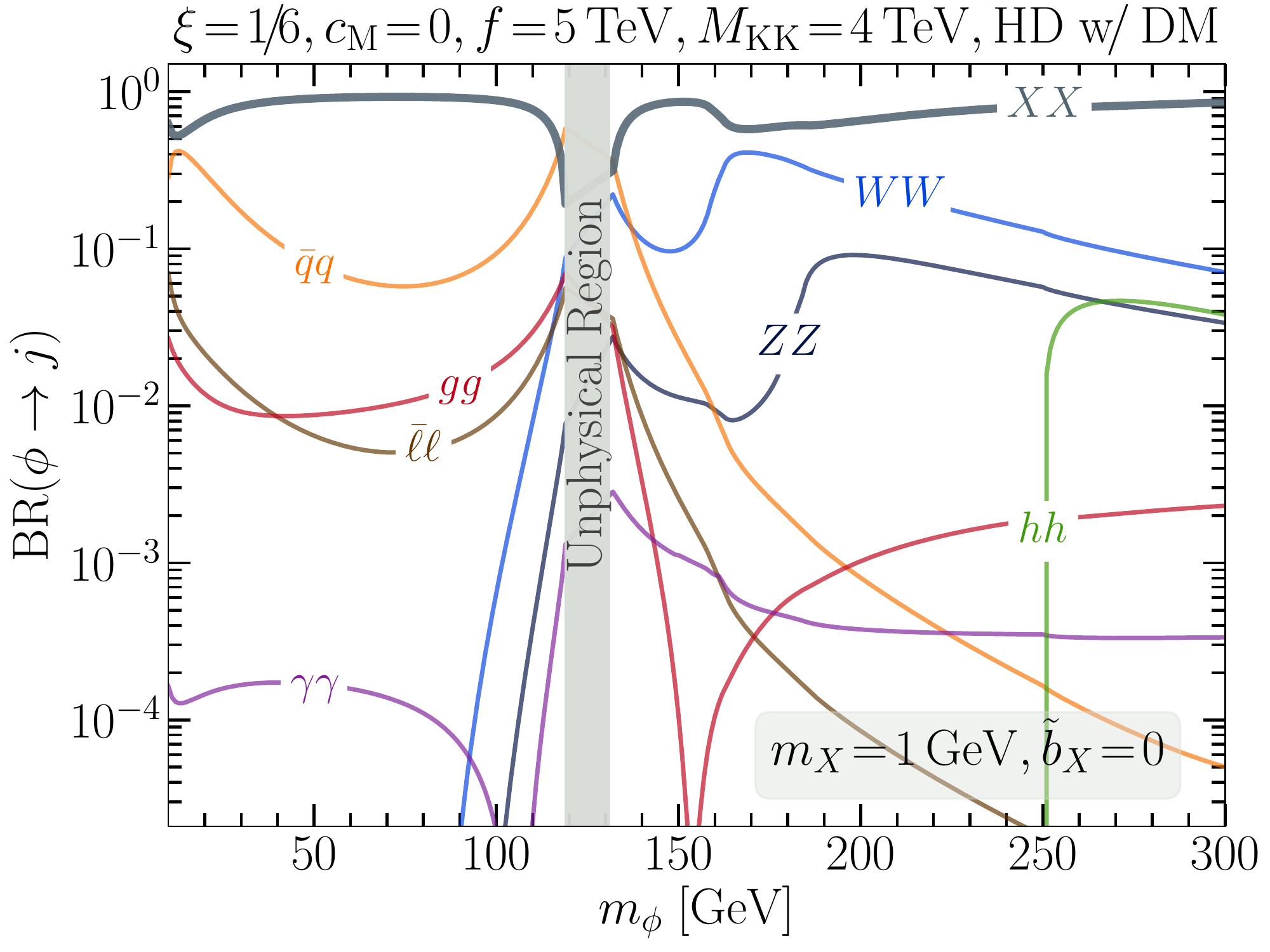} 
\caption{Branching fractions of the dilaton in the holographic dilaton scenario as a function of the dilaton mass $m_\phi$. The upper (lower) panels are for $\xi\!=\!0 (1\!/\!6)$ without (left-panel) and with (right-panel) the presence of dark vector dark portal, respectively. The gray vertical band in the lower-panel plots corresponds to the unphysical region for the given parameters.}
\label{fig:bRSrs}
\end{figure}

\paragraph{Branching ratios:}
In Fig.~\ref{fig:bRSrs}, we show the branching fractions of the dilaton in the holographic dilaton case without (left-panel) and with (right-panel) the presence of the dark sector. 
We remind that the dark sector is constituted by a dark photon which couples to the dilaton through a term proportional to the mass (set by $\cx$) and with a term proportional to the $\beta$-function coefficient 
of the dark photon (set by $\bx$). In the holographic dilaton model we take $\cx=1$ and $\bx=0$ and we assume the dark vector mass to be $m_X\!=\!1\gev$ for concreteness.

The left upper panels of Fig.~\ref{fig:bRSrs} corresponds to no dilaton-Higgs kinetic or mass mixing scenario, and without the presence of dark portal coupling.
This case is very similar to the minimal dilaton model studied in the previous section. 
The main difference is in the gluon gluon decay mode which is reduced due to reduction 
in the value of the $\beta$-function coefficients.
A consequence of this is actually a slight increase in the di-photon branching ratio, which is visible by comparing Fig.~\ref{fig:bRSrs} and Fig.~\ref{fig:mm_brs}. The other mild difference is in a small reduction of the coupling with the massive gauge bosons (see Tab.~\ref{tab:coeff_models}). 

The upper-right panel of Fig.~\ref{fig:bRSrs} corresponds to the case of no dilaton-Higgs mixing but in the presence of the dark sector portal. In this case, as in the minimal dilaton scenario, the dominant branching fraction is to dark photons in the mass range $R_1,R_2$ and it gets comparable to massive SM gauge bosons in the mass range $R_3$. 

In the lower-panels of~Fig.~\ref{fig:bRSrs}, we instead consider non-minimal kinetic mixing between the Higgs and dilaton at the conformal point, i.e. $\xi\!=\!1/6$ or $\bk\!=\!v/f$. 
At the conformal point the dilaton does not couple to massive fields in the trace of energy-momentum tensor. 
Nevertheless, as it can be seen in the plots, the dilaton has considerable branching fractions to the massive fields. 
The reason for such couplings are twofold: (i) such couplings are induced due to the dilaton-Higgs mixing proportional to $\sin\theta$ and they are stronger for dilaton mass closer the Higgs mass $125\gev$, and (ii) the dilaton has non-zero coupling to massive gauge bosons due to the fact that the gauge fields reside in the bulk.
Note that in the dilaton mass regions $R_1$ and $R_2$, in the absence of dark portal (lower-left panel) the dominant branching fraction is to bottom quarks, while in the presence of the dark portal (lower-right panel) the dominant fraction is to the dark vector $X_\mu$. 
Furthermore, there is a sharp dip in the $gg$ channel for $\mphi\!\approx\!155\gev$, which is due to accidental cancelation of the top loop contributions proportional to $F_{1/2}(\tau_t)$, 
the anomalous gauge contribution $b_3$, and the bulk gauge contribution proportional to $1/(kR)$. 
At the same time, for higher dilaton masses $m_\phi\gtrsim 2m_W$ the branching fractions to dark sector are comparable to the massive gauge bosons in the left and right-panels, respectively. The grey band around dilaton mass $\sim\!125 \gev$ represents the unphysical region where the correct Higgs and dilaton masses can not be obtained from the model parameters. 

\begin{figure} [t!]
\centering
\includegraphics[width=0.49\textwidth]{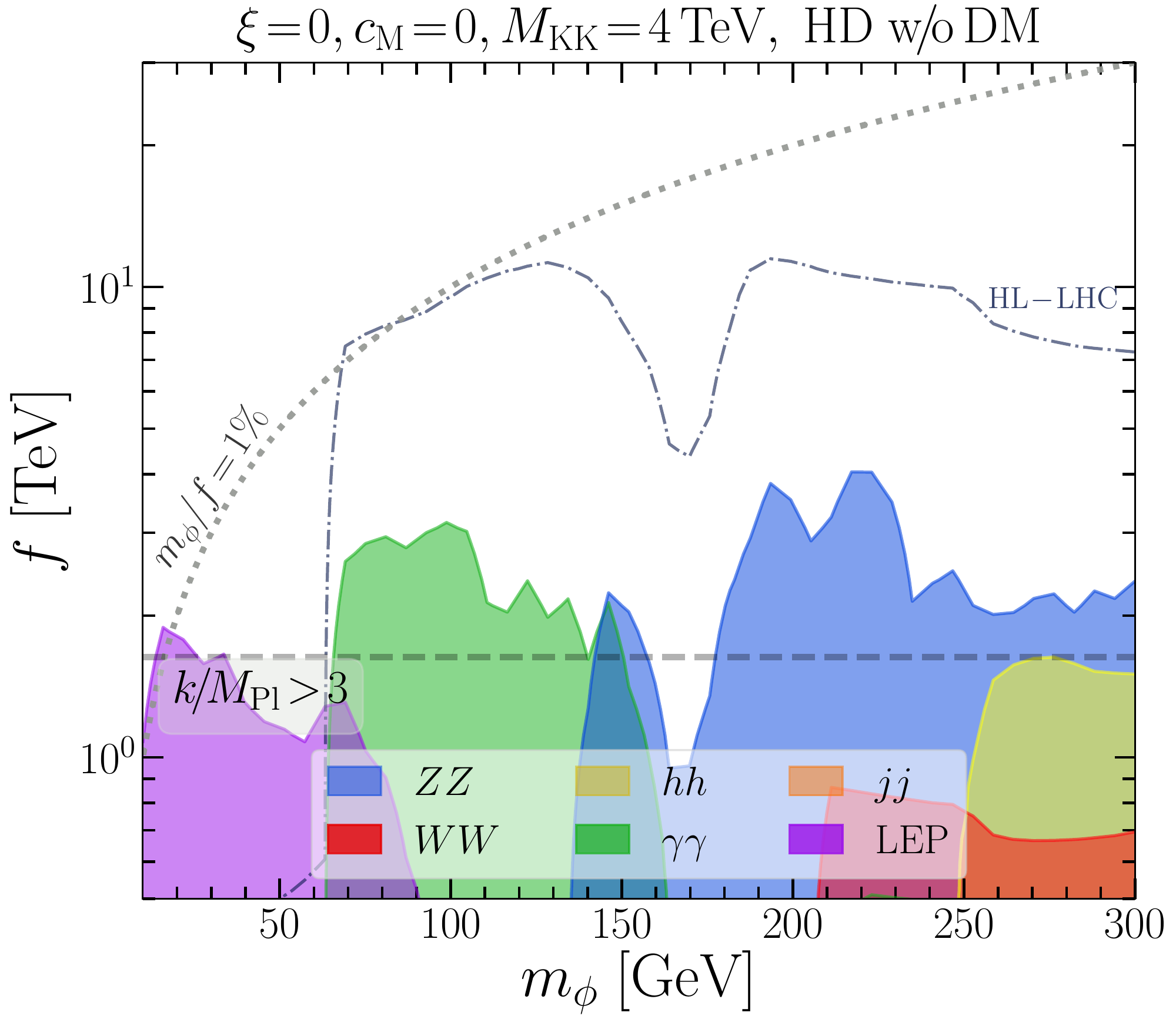}\hspace{-5pt}
\includegraphics[width=0.49\textwidth]{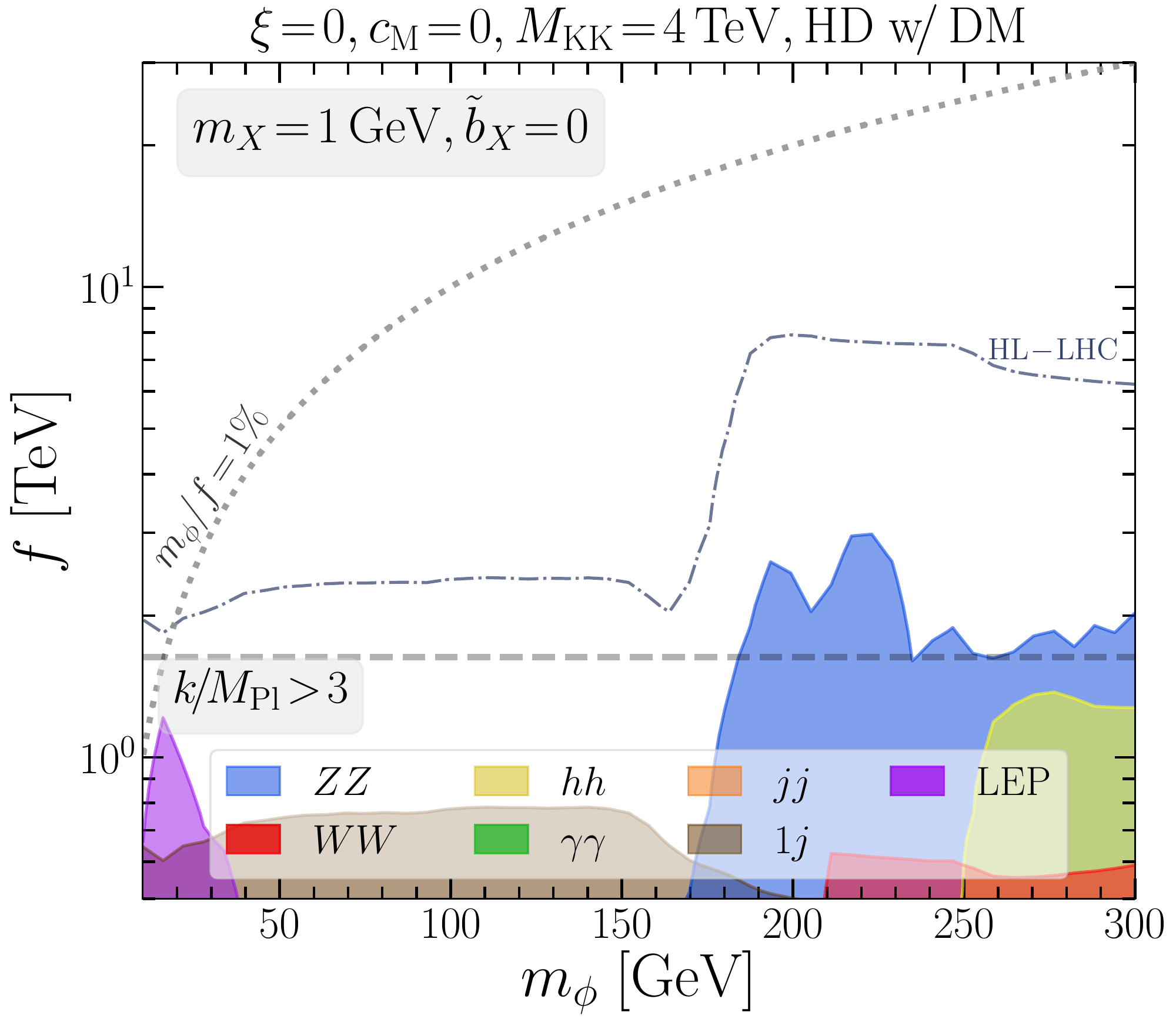}\\
\includegraphics[width=0.49\textwidth]{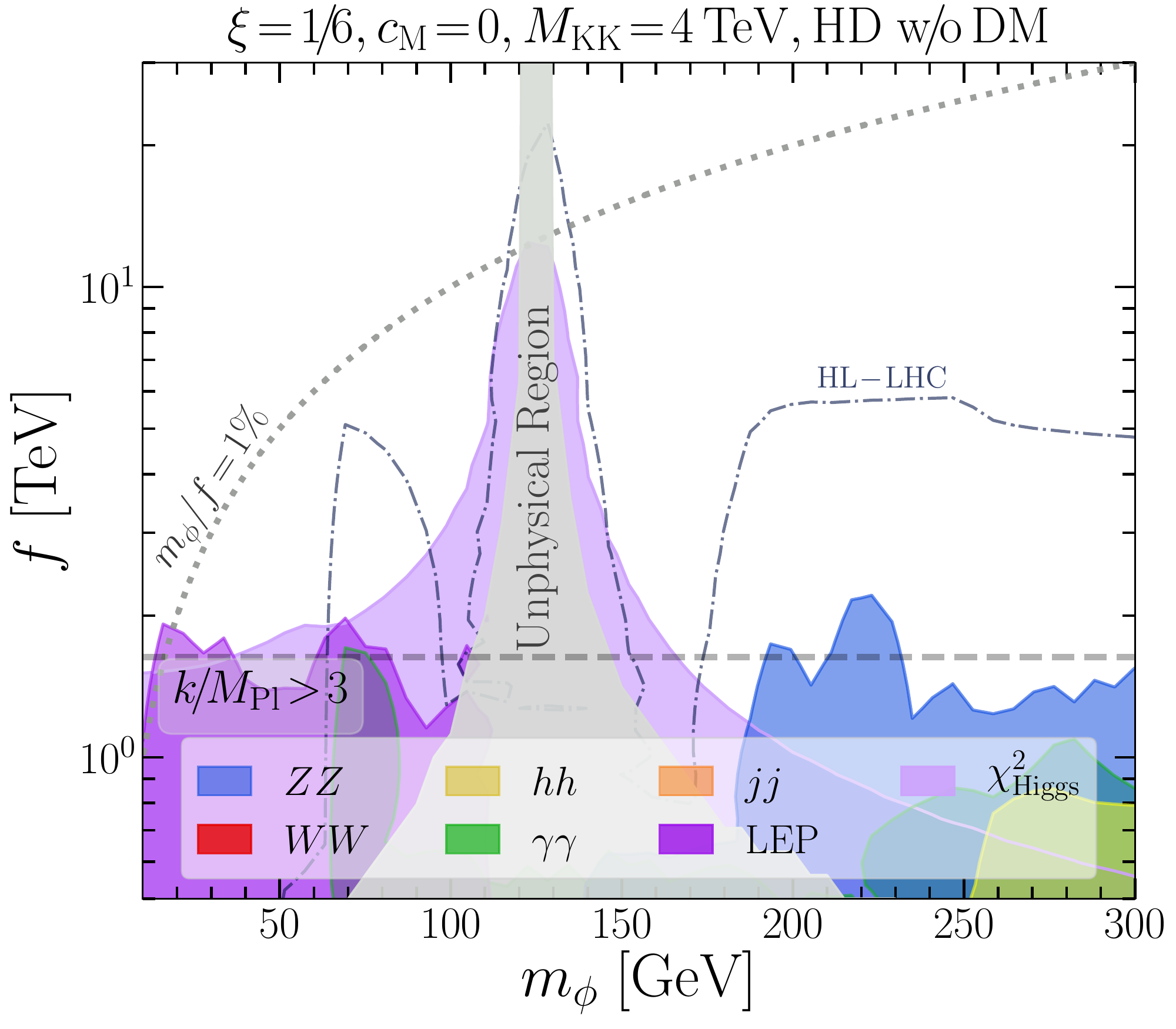} \hspace{-5pt}
\includegraphics[width=0.49\textwidth]{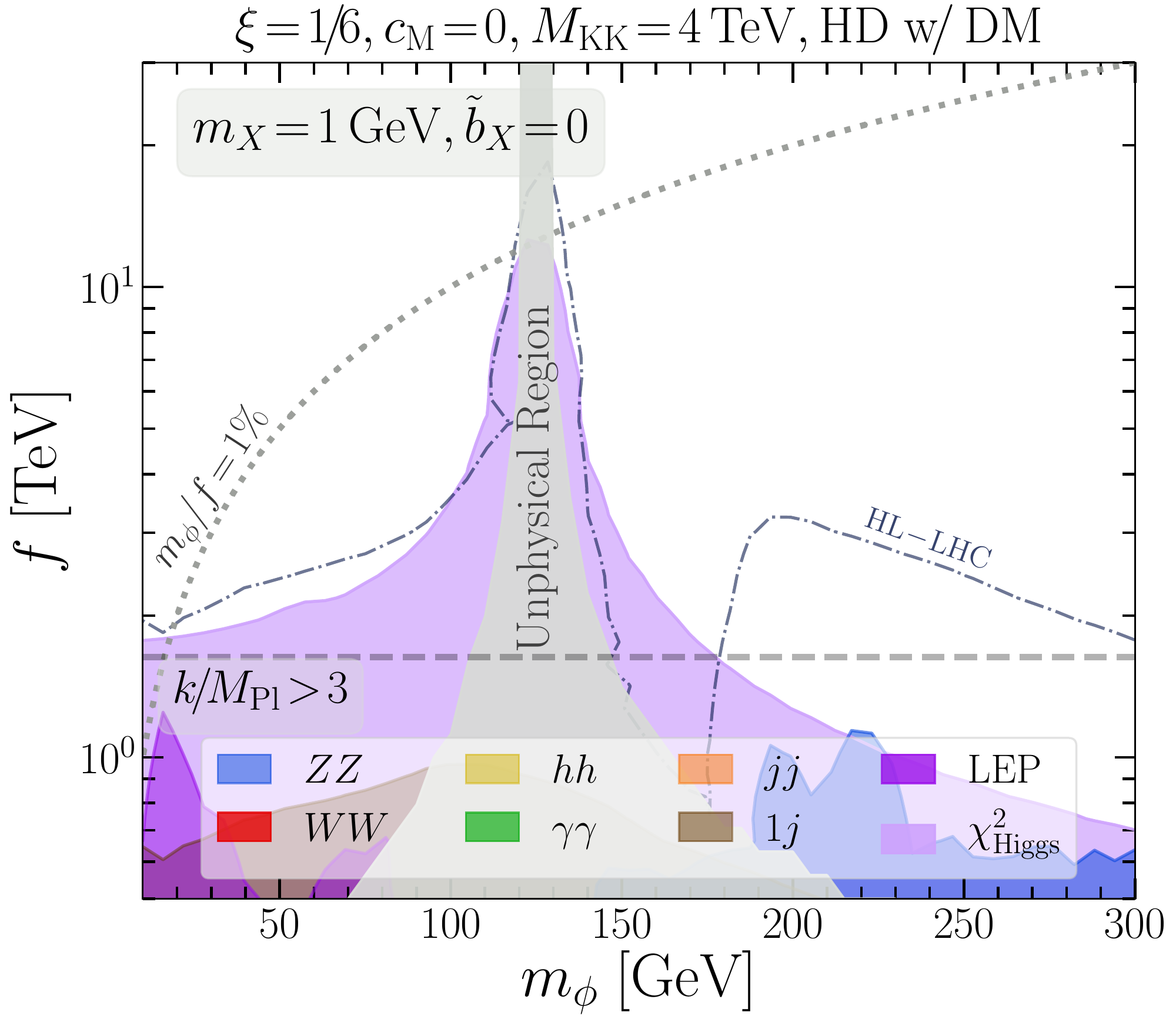} 
\caption{Exclusion bound on the holographic dilaton parameter space in the $m_\phi$ vs $f$ plane. The gray region in the lower-panel plots corresponds to the unphysical region.}
\label{fig:bRS}
\end{figure}

\paragraph{Collider constraints:}
In Fig.~\ref{fig:bRS} we present the current and future collider reach on the parameter space  $m_\phi$ vs $f$ in the holographic dilaton case, without (left-panels) and with (right-panels) dark sector portal.
The upper-panels of Fig.~\ref{fig:bRS} show the case for no dilaton-Higgs mixing, i.e. $\bk\!=\!0$ ($\xi\!=\!0$), while the lower-panels of Fig.~\ref{fig:bRS} present the case with a dilaton-Higgs kinetic mixing at the conformal point, i.e. $\bk\!=\!v/f$ ($\xi\!=\!1/6$).

The shaded areas are excluded by direct LHC searches, LEP constraints, or Higgs coupling fit, as explained in the legend.
We also draw an horizontal dashed grey line marking the validity region for the effective theory, that is where $k/M_{\text{Pl}} \simeq 3$.
Values of $f$ smaller than that line should be considered not theoretically valid with a fixed KK scale at 4 TeV.
Alternatively, one can argue that in such region of parameter space, in order for the effective theory to be valid, the KK scale should be lower,
and hence complementary direct LHC searches for KK gluons and/or KK gravitons should become relevant.

As mentioned, the case with vanishing kinetic and mass mixings $\bk\!=\!0$ and $\cm =0$ is very similar to the minimal model with no mixings, however with one important difference of the reduction in the gluon couplings in the holographic case. 
Phenomenologically this difference is very important as the production cross-section of the dilaton at the LHC is due to gluon fusion. 
The effective dilaton-gluon couplings defined in~\eqref{eq:Cphigg} are different in the two cases due to different $b_3$ coefficient. 
Note that in the holographic case the $b_3\!=\!-1/3$ coefficient contribution is only from the right-handed top quark $t_R$ which is fully composite. 
The cross section of the dilaton via gluon fusion is hence reduced from the minimal dilaton case ($b_3\!=\!7$) to the holographic dilaton ($b_3\!=\!-1/3$) by the following quantity,
\beq
\frac{{\cal C}^{2,{\rm HD}}_{\phi gg}}{{\cal C}^{2,{\rm MD}}_{\phi gg}} \simeq \frac{\big\vert (g_\phi+ \tilde g_\phi)F_{1\!/\!2}(\tau_t) -2(-\frac13+\frac{2}{\alpha_s kR})\tilde g_\phi\big\vert^2}{\big\vert (g_\phi+ \tilde g_\phi)F_{1\!/\!2}(\tau_t) -14\,\tilde g_\phi\big\vert^2},		\label{eq:CphiggAB}
\eeq
where once again we included only the contribution from the top loop.
For instance, in the absence of dilaton-Higgs mixing, i.e. $\bk\!=\!0$, and by setting $kR\!\approx\!10$, the above ratio is $\sim 1/10$. 
Hence, the constraints on $f$ reduces of about a factor of $\sim\!3$ in the holographic model as compared to minimal dilaton (for the same cross section).
The weakening of the constraints on the scale $f$ are indeed manifestly visible in Fig.~\ref{fig:bRS}.
In the case of no dilaton-Higgs mixing and no dilaton-dark sector interaction, we obtain the bound $f\!\gtrsim\!2\tev$.
The presence of the dark portal weakens even further the constraints on $f$, as it is visible in the up right plot of Fig.~\ref{fig:bRS}.
This is particularly significant in the low mass region. The reason is that in the high mass region, the introduction of the dark sector portal only slightly reduces
the branching ratios into massive gauge bosons. Instead, in the low mass region, the presence of the large invisible decay significantly suppress the di-photon branching ratio and this 
reduce considerably the di-photon bound. At the same time, the limits from mono-jet are not strong enough to significantly constrain the parameter space.
In this scenario there are significant portions of parameter space beyond the line $k/\mpl \!\sim\! 3$ which are still experimentally allowed. 
As mentioned, in such regions
direct resonant searches of the KK-modes are naturally complementary strategies to probe the BSM theory.

The lower-left (-right) panel of Fig.~\ref{fig:bRS} corresponds to conformal mixing~$\xi\!=\!1\!/\!6$ between the Higgs and dilaton without (with) the dilaton-dark sector interactions. 
As discussed above, the branching fractions of the dilaton to the massive SM fields are reduced at the conformal point, 
which leads to reduced limits on the scale $f$ in the lower-panel of Fig.~\ref{fig:bRS}. 
Furthermore, at the conformal point $\xi\!=\!1/6$ the effective couplings of the dilaton with gluons parameterized in Eq.~\eqref{eq:Cphigg} has an accidental suppression at $m_\phi\!\approx\!155\gev$ for the $b_3\!=\!-1/3$~\eqref{eq:caseB}. 
This leads to sharp drop of the dilaton production cross section at the LHC for $m_\phi\!\approx\!155\gev$, weakening the LHC bounds. However, this mass region is partly unphysical, the grey region, and partly it is strongly constrained by the Higgs coupling fit, simply because the dilaton-Higgs mixing becomes large for masses of the dilaton closer to the Higgs mass.

In the lower-right plot of Fig.~\ref{fig:bRS}, we see that the presence of the dilaton-dark sector portal weakens the constraints on the interaction scale $f$ derived from the dilaton decay into SM states.
At the same time, the model is now probed by mono-jet searches, potentially relevant in the low mass dilaton region $R_1$.
However, we observe that this case is dominantly constrained by the global $\chi^2$ fit to the Higgs data, which can bound the scale $f$ to be $\sim\!4\tev$ for dilaton mass $m_\phi\!\lesssim\!2m_W$, and it is essentially independent of the dilaton-dark sector portal interaction.
We conclude that for the holographic dilaton the interesting region for the dilaton mass (i.e. $m_{\phi}/f > 1\%$) will be only partially covered also considering HL-LHC, 
as a consequence of the reduced gluon fusion production cross section.

\subsection{Gauge-philic dilaton}

The gauge-philic dilaton case is the simplest in its disguise and involves a minimal set of free parameters including $m_\phi$, $f$, and the gauge-philic gauge couplings $b_i$'s. 
We assume the values of the $b_i$-coefficients are the same as in the SM, i.e. $b_3\!=\!7$, $b_2\!=\!19/6$, and $b_1\!=\!-41/6$. 
Regarding the dark sector portal parameter, we take the dark vector mass $m_X\!=\!1\gev$ and consider a non-zero $\bx$ (the $\beta$-function coefficient of the dark $U(1)_X$ gauge coupling),
while setting to zero the coupling proportional to the dark vector mass, i.e. $\cx=0$.
We remain agnostic about the exact field contents in the dark sector, therefore we do not have exact value of $\bx$. 
Furthermore, the dark gauge coupling $\alpha_\textsc{x}\!\equiv\! g_\textsc{x}^2/4\pi$ is also an unknown parameter. 
On the other hand, what matters for the phenomenology of the dilaton is the combination $\tilde b_\textsc{x}\!$ defined as,
\beq
\tilde b_\textsc{x}\equiv \frac{\alpha_\textsc{x}}{8\pi}\bx.		\label{eq:bxtilde}
\eeq
Hence, as a benchmark scenario we consider the value $\tilde b_\textsc{x}\!=\!0.01$ which could correspond to a dark coupling of order $1$ with a beta coefficient of order a few.

\begin{figure} [t!]
\centering
\includegraphics[width=0.5\textwidth]{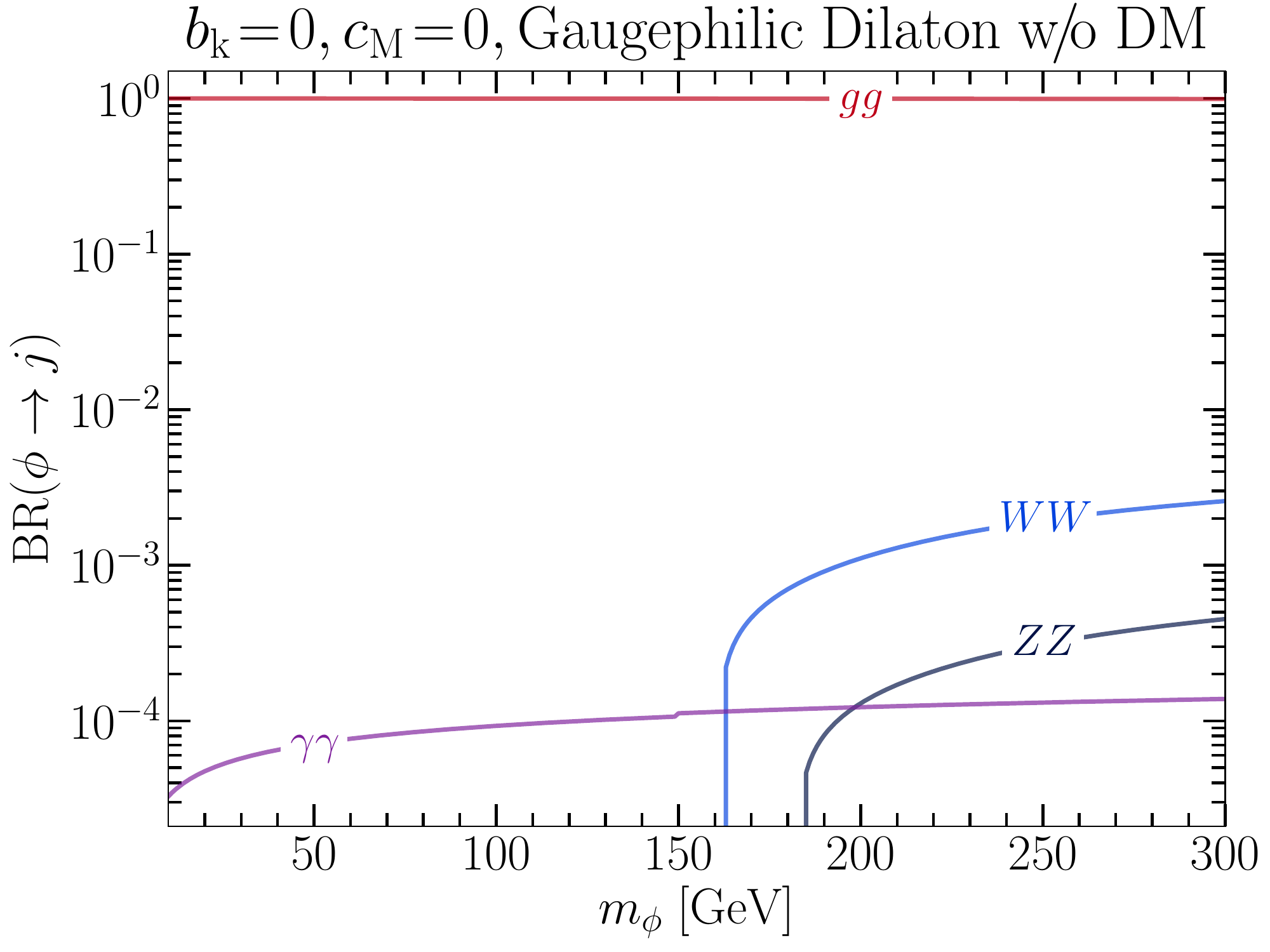}\hspace{-5pt}
\includegraphics[width=0.5\textwidth]{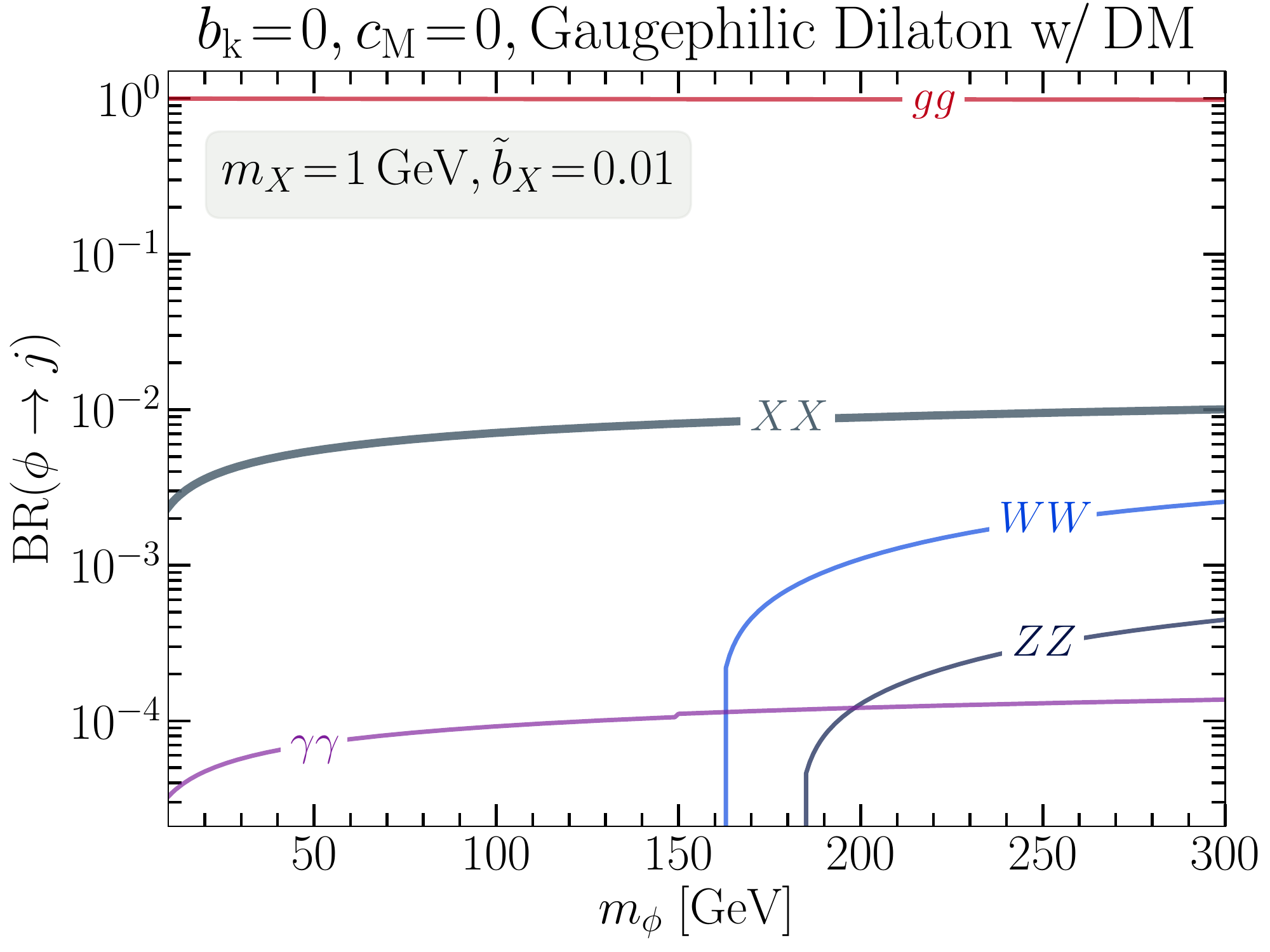}
\caption{Branching fractions of the dilaton in the gauge-philic dilaton scenario as a function of dilaton mass $m_\phi$ for different choices regarding the vector DM.}
\label{fig:anom_brs}
\end{figure}

\paragraph{Branching ratios:}
In Fig.~\ref{fig:anom_brs}, we show the branching fractions of the dilaton to the gauge bosons, without (left-panel) and with (right-panel) the dark portal interaction. 
The gluon channel is the most dominant, whereas the di-photon branching fraction is about $10^{-4}$ times smaller than the gluon fraction. This can understood as the difference of the strength of the coupling constants and $b$-coefficients of the two gauge fields, i.e.
\beq
\frac{\Gamma^\phi_{\gamma\gamma}}{\Gamma^\phi_{gg}}=\frac{1}{8}\bigg\vert\frac{\alpha_\textsc{em}b_\textsc{em}}{\alpha_{s}b_3}\bigg\vert^2\approx 10^{-4}.
\eeq
For the mass of dilaton above $2m_W$, the $WW$ and $ZZ$ channels become active and their branching fractions are of the order $\sim\!10^{-3}$.
The right-plot of Fig.~\ref{fig:anom_brs} shows the case when $\tilde b_\textsc{x}\!=\!0.01$, the branching fraction to vector DM is about $1\%$. In the limit $m_\phi\gg m_X$ (and $\cx=0$), the dilaton partial width to dark vectors is~\eqref{eq:GamX},
\beq
\Gamma^{\phi}_{XX}=\frac{m_{h,\phi}^3}{4 \pi  f^2}\tilde b_\textsc{x}^2\,.
\eeq
Therefore, it is straightforward to generalized the result of Fig.~\ref{fig:anom_brs} for different $\tilde b_\textsc{x}$ values.
Note the branching fraction is independent of the interaction scale $f$. 

\begin{figure} [t!]
\centering
\includegraphics[width=0.5\textwidth]{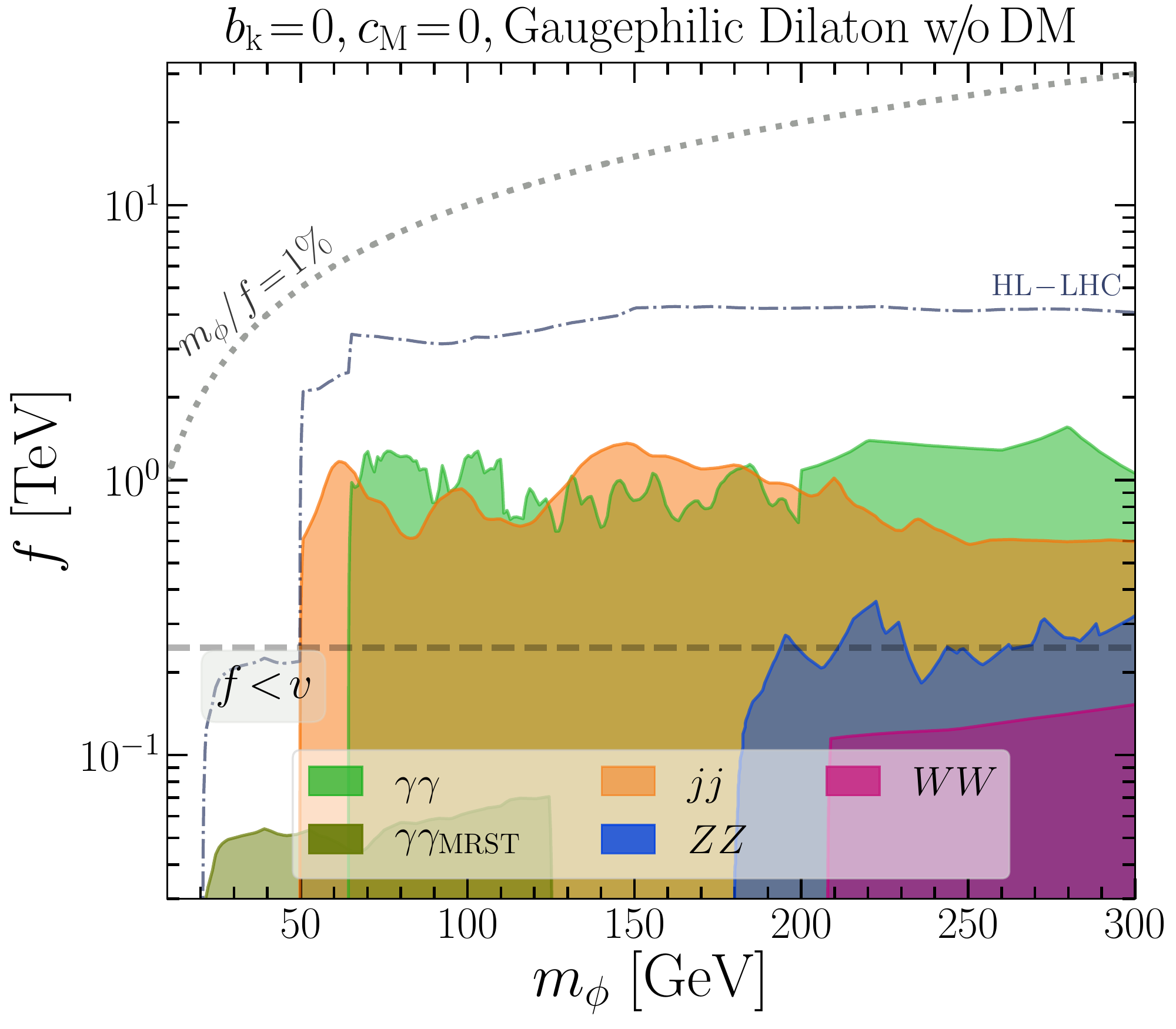}\hspace{-5pt}
\includegraphics[width=0.5\textwidth]{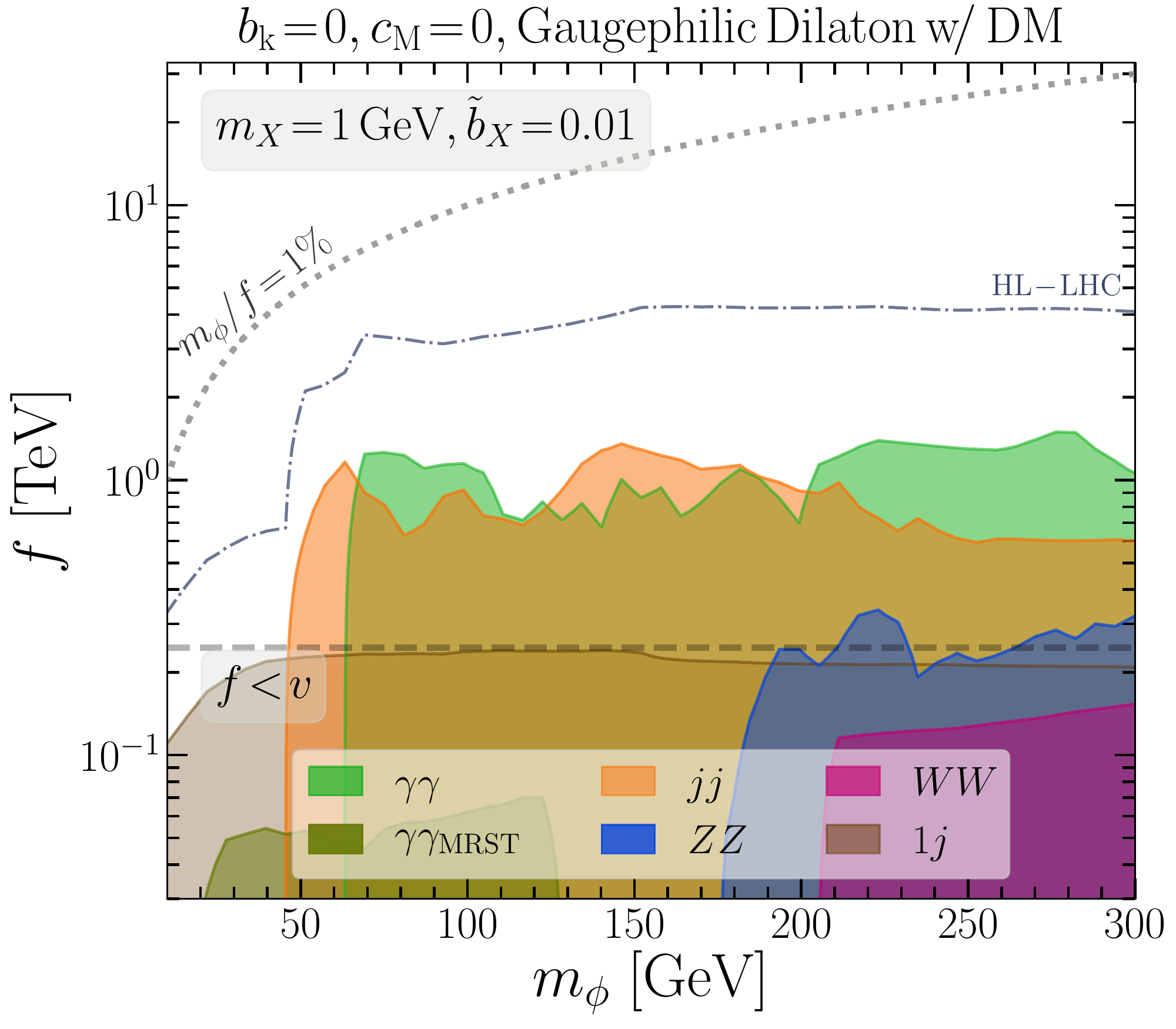}
\caption{Exclusion bound on the gauge-philic dilaton parameter space in the $m_\phi\!-\!f$ plane.}
\label{fig:anomdil}
\end{figure}

\paragraph{Collider constraints:}
In the left and right panel of Fig.~\ref{fig:anomdil}, we present the allowed parameter space of the gauge-philic dilaton scenario. 
The shaded regions are excluded by different searches coded in color. 
Note that for the left-plot (no dark portal), the dominant constraint are from the di-photon and di-jet resonant searches.
However, these constraint start from $50\gev$ and the lower mass region is pretty much unconstrained. 
For the dilaton mass $\mphi\gtrsim50\gev$ the di-photon and di-jet searches constraint $f\sim1\tev$ with the run-2 data. However with the HL-LHC these constraint can improve by a factor $\sim\!4$ (dashed-dotted curve).
In the low mass range we overlaid the constraint due to di-photon cross section measurement, $\gamma\gamma_{\rm MRST}$, as derived in Ref.~\cite{Mariotti:2017vtv}.
However, overall the parameter space is less constrained in the low mass region $m_\phi\!\leq\!50\gev$.

We then consider the case when dark portal interaction is present (right panel in Fig.~\ref{fig:anomdil}).
In the low mass region ($m_\phi\!\leq\!50\gev$) there is some additional coverage due to the mono-jet searches. 
For heavier dilaton, i.e. $\mphi\gtrsim50\gev$, the di-photon and di-jet still dominate and the limits are similar to the scenario without the dark portal (left-plot). 
Note that for larger values of  $\tilde b_\textsc{x}$ the mono-jet constraint becomes very significant as the branching ratio scales as $\tilde b_\textsc{x}^2$.
However, values of the $\tilde b_\textsc{x}$ coefficient larger than $1\%$ require large dark $U(1)_X$ beta function coefficient $\bx$ and/or strong coupling regime, i.e. $g_\textsc{x}\gtrsim1$.

Finally, the dashed horizontal line in Fig.~\ref{fig:anomdil} indicates the region where $f < v$ and hence where our effective theory description is not valid anymore.
We nevertheless show the collider reach in such region for illustrative purpose in order to highlight the significant opportunity for improvements in experimental coverage of such scenario.
We conclude by observing that in this simplified gauge-philic scenario there is a large portion of parameter space with dilaton mass $m_\phi\!/\!f\!>\!1\%$ which is still viable and 
that could be (at least partially) covered at HL-LHC.

\section{Conclusions}
\label{sec:con}

In this paper we studied the collider phenomenology of a light dilaton, focusing on the dilaton mass range $[10-300]$ GeV. 
The effective theory includes the SM plus a light scalar (the dilaton) coupled through higher dimensional operators to the SM, suppressed by the scale $f$ of spontaneous breaking of the scale invariance. 
We also included possible sources of mixing (kinetic or mass) between the dilaton and the SM-Higgs.
In addition, we consider the case in which the dilaton acts as a portal to a dark sector which respects the non-linearly realized scale symmetry. 
The lightest state in the dark sector is a dark gauge boson of a $U(1)_X$ gauge symmetry, with mass much smaller than the dilaton mass. 
This allows the dilaton to decay invisibly and leads to missing energy signatures at the LHC.

We considered three benchmark scenarios with different values for the effective couplings, that can be mapped to possible UV completions.
First, we consider a minimal dilaton model where the dilaton and the SM Higgs can have a mass mixing. 
Second, we investigate the case of a holographic dilaton (within the paradigm of partial compositeness) and where we introduced also a dilaton-Higgs kinetic mixing.
The third scenario we consider is then a gauge-philic dilaton where the dilaton has only couplings to the gauge bosons via the running of gauge couplings, and that results elusive for collider searches.

We explore in detail the parameter space in the $m_{\phi}$ vs $f$ plane for these benchmark scenarios with the available LHC analysis 
and we point out the accessible region for future HL-LHC.
The interesting conclusion is that in minimal dilaton case where the coupling of the dilaton with gluons is determined by the full SM $\beta$-function coefficient, the entire region with $m_{\phi}/f > 1\%$ will essentially be covered at the HL-LHC. This is valid even if the dilaton acts as a portal to a dark sector and has a sizeable branching fraction into missing energy, because of the impact of the mono-jet constraints.

In the holographic dilaton model, where we assumed only the Higgs doublet and the right-handed top quark as composite states, the dilaton coupling to gluons is reduced roughly by $1/3$ as compared to the minimal dilaton case. 
This implies a weakening of the constraints on $f$ by a factor of $3$ in the holographic dilaton case. Hence, large region of the parameter space with $m_{\phi}/f >1\%$ is still allowed with present LHC data and will be only partially covered by the future HL-LHC. 
Similar features appears also in the phenomenology of the gauge-philic dilaton. In that case the coupling with gluons is sizeable but the coupling with photons and massive gauge bosons, which normally drive the
LHC limits, are reduced compared to the minimal dilaton case.

A common conclusion for all the benchmarks is that the very low mass region $m_{\phi} \!<\! 60\gev$ remains still poorly covered and suggests that dedicated LHC analysis could improve the reach for such low masses.
In general, the region with $m_{\phi}/f >10\%$ is already significantly constrained by LHC searches and will be completely covered at the HL-LHC, while the $m_{\phi}/f >1\%$ will be at least partially probed by HL-LHC.
We conclude that there is a sizeable and interesting portion of parameter space, without a large hierarchy between $m_{\phi}$ and $f$, where the future runs of the LHC can look for the existence of a light dilaton.

\section*{Acknowledgements}
We would like to thank Riccardo Argurio, Brando Bellazzini, Zackaria Chacko and Diego Redigolo for discussions.
This work is supported by FWO under the EOS-be.h project no. 30820817 and Vrije Universiteit Brussel through the Strategic Research Program ``High Energy Physics''.

\appendix

\section{Feynman rules and useful formulae}
\label{sec:feyn_rules}
For completeness, in this Appendix we collect all the relevant Feynman rules, partial widths and formulae employed in the main text. The Feynman rules are given in a compact form in Tab.~\ref{fig:feynrules}. 
It is straightforward to calculate the partial widths with the given Feynman rules. 
\begin{table}[t!]
\centering
\includegraphics[width=0.97\textwidth]{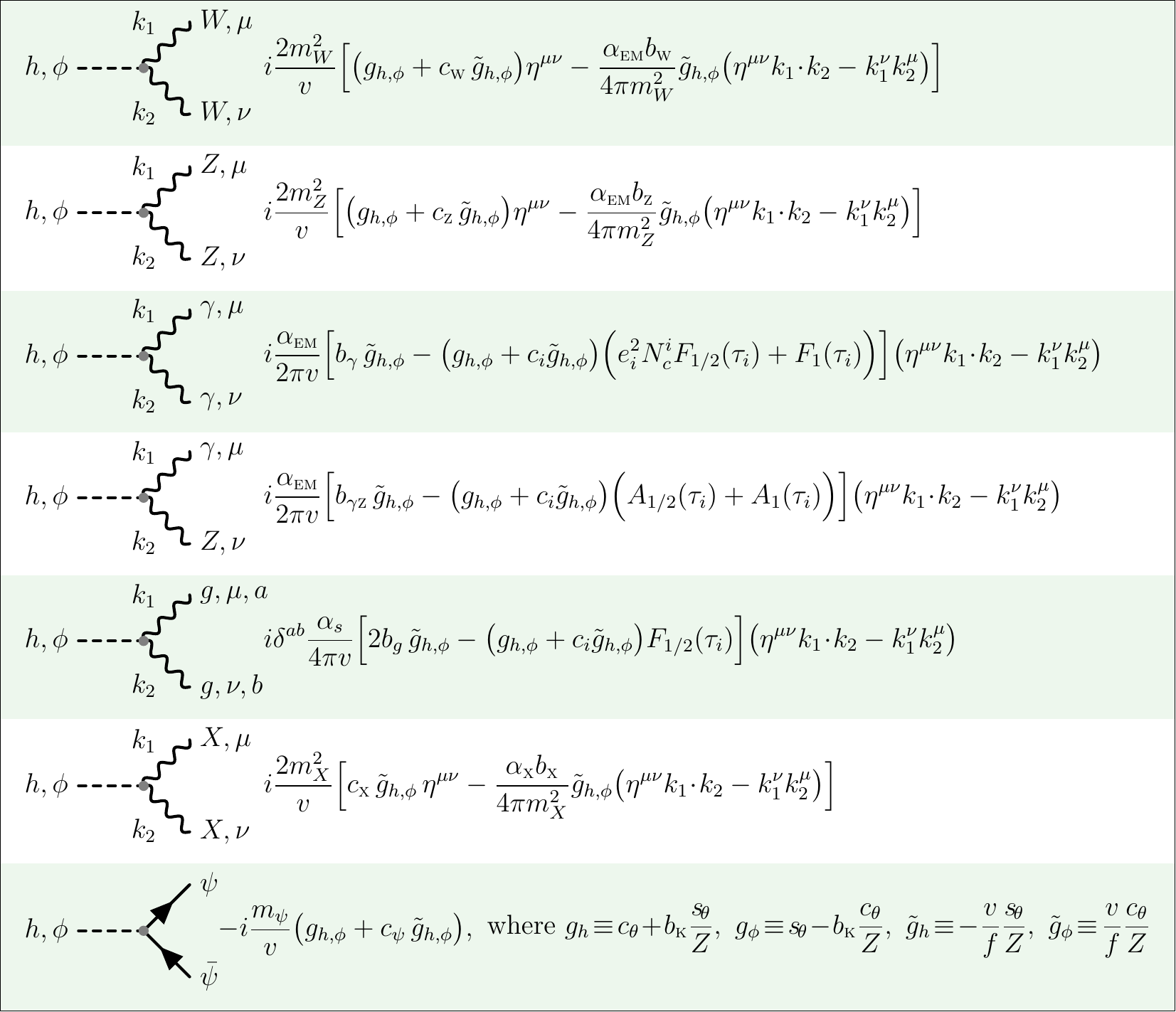}
\caption{Feynman rules for the SM-like Higgs $h$ and the dilaton $\phi$ couplings with the SM particles. The triangle loop functions $F_{s}$ and $A_{s}$ are collected below where the subscript denote the spin of the particle in the loop and the sum over all the particles in the corresponding loops is understood. The model dependent coefficients $b$'s and $c$'s can be found in Tab.~\ref{tab:coeff_models}.}
\label{fig:feynrules}
\end{table}

\subsubsection*{Trilinear scalar interactions} 
As discussed in Sec.~\ref{sec:theory}, the trilinear Higgs couplings are more involved in our case in the presence of dilaton-Higgs mixing. In particular, when the mass of dilaton is larger (smaller) than twice (half) of the SM Higgs mass than the dilaton (Higgs) would decay to two Higgs (dilaton) states. The partial width of a heavy state $i$ decaying to a lighter state $j$ is given by, 
\beq
\Gamma(i\!\to\! jj)=\frac{g_{ijj}^2}{32\pi}\frac{1}{m_i}\sqrt{1-\frac{4m_j^2}{m_i^2}}, 
\eeq
where $i/j$ can be the dilaton or Higgs depending on the dilaton mass. The trilinear couplings $g_{ijj}$ are derived from all the sources of trilinear interactions discussed in Sec.~\ref{sec:theory}.
After applying the rotational matrix~\eqref{h-r_mixing}, in the physical mass eigenbasis ($\phi,h$), these trilinear couplings have the form,
\begin{align}
g_{h\phi\phi}&= \frac{1}{f Z^3}(s_{\!\theta} Z-\bk  c_{\theta}) \left(3 \bk  c_{\theta} s_{\!\theta}+2 c_{\theta}^2 Z-s_{\!\theta}^2 Z\right) \left(2
   \bh  m_{h}^2-\bh  m_{\phi}^2-4 \ch  m_{h_0}^2\right)	\notag\\
   &\qquad	-\frac{3 m_{h_0}^2}{v Z^3} (\bk  s_{\!\theta}+c_{\theta}
   Z) (\bk  c_{\theta}-s_{\!\theta} Z)^2\,,	\label{eq:ghpp} \\
g_{\phi hh}&=
\frac{1}{f Z^3}(c_{\theta} Z+\bk  s_{\!\theta})\left(3 \bk  c_{\theta} s_{\!\theta}+c_{\theta}^2Z-2 s_{\!\theta}^2Z\right) 
   \left(2 \bh  m_{h}^2-\bh  m_{\phi}^2-4 \ch  m_{h_0}^2\right)	\notag\\
   &\qquad+\frac{3 m_{h_0}^2}{v Z^3} (\bk 
   c_{\theta}-s_{\!\theta} Z) (\bk  s_{\!\theta}+c_{\theta} Z)^2\,.
   \label{eq:gphh}
\end{align}
Note that in the absence of the dilaton-Higgs kinetic and mass mixings, i.e. $\bk\!=\!0$ and $\cm\!=\!0$ (hence $s_{\!\theta}\!=\!0$, $Z\!=\!1$, $\bh\!=\!1$, $\ch\!=\!1$), the above trilinear couplings simplify to
\beq
g_{h\phi\phi}=0, \qquad g_{\phi hh}=-\frac{\mphi^2}{f}\Big(1+\frac{2 m_h^2}{\mphi^2}\Big).
\eeq
Whereas, in the case of minimal dilaton, where $\bk\!=\!0$ and $\cm\!\neq\!0$ (hence $s_{\!\theta}\!\neq\!0$, $Z\!=\!1$, $\bh\!=\!1$, $\ch\!=\!1$), we get the trilinear couplings as, 
\begin{align}
g_{h\phi\phi}&=\frac{s_{\!\theta}}{f}\Big[ \left(2c_{\!\theta}^2-s_{\theta}^2\right) \left(2m_h^2-4 m_{h_0}^2- m_{\phi
   }^2\right)\Big]-\frac{3c_{\theta } s_{\!\theta}^2}{v}   m_{h_0}^2\,	,	\label{eq:ghpp_mm} \\
g_{\phi h h}&=\frac{c_{\theta }}{f}\Big[\left(c_{\theta}^2-2
   s_{\!\theta}^2\right)\left(2m_h^2-4 m_{h_0}^2- m_{\phi
   }^2\right)\Big] - \frac{3c_{\theta }^2 s_{\!\theta}}{v} m_{h_0}^2\,.	\label{eq:gphh_mm}
\end{align}

The SM trilinear coupling $g_{hhh}$ can be an important probe of SM Higgs mixing with other scalars. In our case, the Higgs trilinear coupling is modified to the following form:  
\begin{align}
g_{hhh}&=\frac{s_{\!\theta}}{2 f Z^3} (\bk  s_{\!\theta}+c_{\theta} Z)^2 \left(\bh m_{\phi}^2-2\bh m_{h}^2+4 \ch 
   m_{h_0}^2\right)-\frac{m_{h_0}^2}{2 v} \left(c_{\theta}+\frac{\bk  s_{\!\theta}}{Z}\right)^3, \label{eq:ghhh}
\end{align}
In the minimal dilaton scenario the above coupling reduces to
\begin{align}
g_{hhh}&=\frac{c_{\theta}^2 s_{\!\theta} }{2 f}\left(4 m_{h_0}^2-2 m_h^2+m_{\phi}^2\right)-\frac{c_{\theta}^3}{2 v} m_{h_0}^2 , 
\end{align}
which obviously in the absence of any mixing reduces to the usual SM value $-m_{h}^2/(2 v) $.
We note that deviations in the Higgs trilinear coupling are not constraining the parameter space we explore. 
However future improved measurement of such coupling could also be used as an indirect probe of the dilaton-Higgs mixing.

\subsubsection*{Dilaton/Higgs partial width to dark photon} 
In the presence of the dilaton-Higgs mixing, both mass eigenstates of dilaton $\phi$ and SM-like Higgs $h$ couple to the dark photon.
The partial width of the Higgs/dilaton to the dark photon $X_\mu$ resulting from the Lagrangian~\eqref{eq:Ldark}.  is given by, 
\beq
\Gamma^{h,\phi}_{XX}\!=\!\frac{m_{h,\phi}^3\tilde g_{h,\phi}^2}{32 \pi  v^2}\! \sqrt{1-\frac{4 m_X^2}{m_{h,\phi }^2}} \!\left[(\cx^2+8 \tilde b_\textsc{x} ^2)-4 (\cx^2+6 \cx\tilde b_\textsc{x} +8 \tilde b_\textsc{x} ^2) \frac{m_X^2}{m_{h,\phi }^2}+12 (\cx+2 \tilde b_\textsc{x} )^2 \frac{m_X^4}{m_{h,\phi }^4}\right]\!,		\label{eq:GamX}
\eeq
where $\tilde g_{h,\phi}$ are given in \eqref{gs_param} and $\tilde b_\textsc{x}$ in Eq.~\eqref{eq:bxtilde}.
Note that in the absence of the coupling $\bx$, the above relation for partial width reduces to the standard width of a scalar to massive gauge bosons, i.e. for $\bx=0$,
\beq
\Gamma^{h,\phi}_{XX}=\frac{\cx^2\, m_{h,\phi}^3\,\tilde g_{h,\phi}^2}{32 \pi  v^2} \sqrt{1-\frac{4 m_X^2}{m_{h,\phi }^2}} \left[1-4 \frac{m_X^2}{m_{h,\phi }^2}+12 \frac{m_X^4}{m_{h,\phi }^4}\right].
\eeq
Note that the partial width~\eqref{eq:GamX} becomes a simple expression in the limit $m_X\ll m_{h,\phi}$,
that is
\beq
\Gamma^{h,\phi}_{XX}=\frac{m_{h,\phi}^3\,\tilde g_{h,\phi}^2}{32 \pi  v^2}\Big(\cx^2+8 \tilde b_\textsc{x} ^2\Big),	\label{eq:Gamphixx}
\eeq
which is independent of the dark photon mass in this limit. 

\subsubsection*{The loop functions} 
The most frequently used form factors are the $F_{1/2} (\tau_i)$ and $F_1(\tau_i)$,  given as
\begin{align}
F_{1/2}(\tau_i)&=-2\tau_i\big[1+(1-\tau_i)f(\tau_i)\big], 
\lsp F_{1}(\tau_i)=2+3\tau_i+3\tau_i(2-\tau_i)f(\tau_i), 
\end{align}
where $\tau_i\equiv4m_i^2/m_\phi^2$, and 
\begin{align}
f(\tau_i)&= \begin{cases}
\arcsin^2\big(1/\sqrt{\tau_i}\big), & \hspace{0.7cm}\text{if }  \tau_i \geq 1,\\    
-\frac14\big[\ln\big(\frac{1+\sqrt{1-\tau_i}}{1-\sqrt{1-\tau_i}}\big)-i\pi\big]^2, & \hspace{0.7cm}\text{if }  \tau_i <1.
\end{cases}
\end{align}
The asymptotic values of these form factors are: 
\begin{align}
F_{1/2}(\tau_i)&= \begin{cases}
-4/3,~~~ &  \tau_i \to \infty\\    
0, &   \tau_i \to 0
\end{cases}, 	&F_1(\tau_i)&= \begin{cases}
7,~~~ &  \tau_i \to \infty\\    
2, &   \tau_i \to 0
\end{cases}.
\end{align}
The exact expressions of the form factors $A_{1/2}(\tau_i)$ and $A_1(\tau_i)$ used in the $Z\gamma$ final state vertex can be found in the Higgs Hunter's Guide~\cite{Gunion:1989we}.


\bibliography{lightDilaton_v2}{}

\end{document}